\pdfoutput=1
\documentclass[10pt, twocolumn, comsoc]{IEEEtran}

\usepackage{graphicx}
\usepackage[noadjust]{cite}
\usepackage{mcite}
\usepackage{amsfonts,helvet}
\usepackage{fancyhdr}
\usepackage{booktabs}
\usepackage{threeparttable}
\usepackage{amsthm}
\usepackage{amsmath}
\usepackage{siunitx}
\usepackage{amssymb}
\usepackage{dsfont}
\usepackage{subfigure}
\usepackage{xcolor}
\usepackage{color}
\usepackage[linesnumbered,ruled,noend]{algorithm2e}
\DontPrintSemicolon 
\usepackage{algpseudocode}
\usepackage{algcompatible}
\usepackage{enumerate}
\usepackage{gensymb}
\usepackage{cancel}
\usepackage{graphicx}
\usepackage{wrapfig}
\usepackage{ragged2e}

\newtheorem{theorem}{Theorem}

\newtheorem{example}{Example}

\newtheorem{lemma}{Lemma}

\newtheorem{proposition}{Proposition}
\newtheorem{remark}{Remark}

\usepackage{bbm}
\usepackage{eucal}

\usepackage{multirow}
\usepackage{dsfont}
\usepackage{boldline}

\usepackage{amssymb}
\usepackage{accents}
\newcommand{\widebreve}[1]{\accentset{\smile}{#1}}

\def\bphi{\pmb{\phi}}

\def\ba{{\bf a}}

\def\bc{{\bf c}}
\def\bd{{\bf d}}
\def\be{{\bf e}}
\def\bff{{\bf f}}

\def\bh{{\bf h}}

\def\bn{{\bf n}}

\def\bq{{\bf q}}
\def\br{{\bf r}}

\def\bt{{\bf t}}

\def\bw{{\bf w}}
\def\bx{{\bf x}}
\def\by{{\bf y}}
\def\bz{{\bf z}}

\def\bA{{\bf A}}
\def\bB{{\bf B}}
\def\bC{{\bf C}}

\def\bF{{\bf F}}

\def\bH{{\bf H}}
\def\bI{{\bf I}}
\def\bJ{{\bf J}}
\def\bK{{\bf K}}
\def\bL{{\bf L}}

\def\bN{{\bf N}}

\def\bQ{{\bf Q}}
\def\bR{{\bf R}}
\def\bS{{\bf S}}

\def\bU{{\bf U}}

\def\bW{{\bf W}}
\def\bX{{\bf X}}
\def\bY{{\bf Y}}
\def\bZ{{\bf Z}}

\def\cC{\mbox{$\mathcal{C}$}}

\def\cF{\mbox{$\mathcal{F}$}}

\def\cL{\mbox{$\mathcal{L}$}}

\def\cN{\mbox{$\mathcal{N}$}}

\def\cQ{\mbox{$\mathcal{Q}$}}

\def\cU{\mbox{$\mathcal{U}$}}

\def\bbC{\mbox{$\mathbb{C}$}}

\def\bbE{\mbox{$\mathbb{E}$}}
\def\bbF{\mbox{$\mathbb{F}$}}

\def\bbH{\mbox{$\mathbb{H}$}}

\def\bbR{\mbox{$\mathbb{R}$}}

\begin{document}
\providecommand{\widebar}[1]{\overline{#1}}

\title{
{A Bussgang-based Latent Lifting Method:
\\
Discrete RIS Channel Estimation and Beamforming}
}
\author{Mintaek~Oh, Namyoon~Lee, and Jinseok~Choi

\thanks{Mintaek Oh and Jinseok Choi are with the School of Electrical Engineering, Korea Advanced Institute of Science and Technology (KAIST), Daejeon 34141, Republic of Korea (e-mail: {\texttt{\{ohmin, jinseok\}@kaist.ac.kr}}). 
}

\thanks{
Namyoon Lee is with the Department of Electrical Engineering, Pohang University of Science and Technology (POSTECH), Pohang 37673, Republic of Korea (e-mail: {\texttt{nylee@postech.ac.kr}}).
}

}

\maketitle \setcounter{page}{1} 

\begin{abstract}
This paper addresses the challenge of discrete phase-shift optimization in reconfigurable intelligent surface (RIS)-aided multiuser communication systems.
Practical RIS hardware often utilizes low-resolution phase shifters, leading to challenging combinatorial optimization problems.
To address this challenge, we propose a Bussgang-based latent lifting (BuLL) method that represents discrete phase shifts as quantized outputs of continuous latent variables.
Our BuLL method calibrates the Bussgang gain and distortion covariance under a reference distribution and keeps these moments fixed during deterministic latent optimization.
These fixed moments are then used to construct a distortion-aware continuous surrogate that incorporates reference-averaged discretization effects without penalty terms.
For uplink channel estimation, we develop a matrix quadratic transform-based algorithm to optimize a mean-square-error  surrogate.
For downlink beamforming, we develop a generalized power iteration-based RIS phase-shift optimization algorithm for a spectral efficiency (SE) surrogate.
Numerical results demonstrate improvements in channel estimation accuracy and sum SE over conventional benchmarks in the evaluated low-resolution RIS settings.
\end{abstract}


\vspace{-0.55cm}
\section{Introduction}
Reconfigurable intelligent surface (RIS) has emerged as a key enabling technology for next-generation wireless communication systems \cite{kumar2025ris}. 
Composed of numerous low-cost passive reflecting elements, RIS can reshape the wireless propagation environment to enhance signal quality and suppress interference without requiring active radio-frequency chains \cite{di2020smart}. 
Consequently, RIS-aided communications have attracted research attention across various applications \cite{wu2019intelligent, kundu2021channel, zheng2022survey}.

While many existing studies assume ideal, continuous (infinite-resolution) phase shifts for theoretical simplicity, hardware implementations impose stringent constraints. 
To maintain low hardware cost and minimal power consumption, practical RIS prototypes are predominantly developed with extremely low-resolution phase shifters, typically 1-bit \cite{pei2021ris} or 2-bit \cite{zhao20242}. 
Therefore, it is  more practical and realistic to design RIS-aided systems under the assumption of discrete phase shifts rather than idealized continuous ones \cite{wu2019beamforming, gao2021robust, ramezani2024joint}.


However, considering discrete phase-shift optimization leads to significant technical challenges.
The discrete nature of the hardware fundamentally transforms the RIS optimization into mixed-integer nonlinear programming (MINLP), which is generally NP-hard \cite{parker2014discrete, burer2012non}. 
In particular, since an exhaustive search over all possible configurations is computationally prohibitive for practical RIS sizes, developing efficient algorithms that effectively handle discrete phase-shift constraints is of paramount importance.

\subsection{Related Works}
Existing approaches for discrete RIS optimization generally fall into three categories: relaxation-based, penalty-based, and combinatorial or heuristic search methods.
The most widely adopted strategy is the relaxation-then-quantization approach \cite{wu2019beamforming_icassp, so2007approximating, di2020hybrid}. 
In this paradigm, discrete phase-shift constraints are relaxed to continuous unit-modulus constraints, solved via techniques such as semidefinite relaxation \cite{gao2021robust}, successive convex approximation \cite{pan2020multicell}, and  manifold optimization \cite{manifold_ref, guo2020weighted, pan2022overview}, then   projected to the nearest discrete values.
While computationally tractable, these methods perform a 
\emph{deterministic relaxation} in which the relaxed continuous variable 
bears no modeled statistical relationship to its eventual discrete value. 
Consequently, the continuous optimization is entirely agnostic to the 
subsequent quantization, leading to performance degradation that is particularly definite in low-resolution regimes where the discretization error is severe.

To mitigate discretization loss, the \textit{penalty-based} approach \cite{zhu2025discrete} augments the objective function to enforce discrete constraints during optimization. 
While yielding improved performance, it suffers from high computational complexity due to double-loop iterations and relies heavily on empirical parameter tuning.
Alternatively, direct discrete optimization relies on \textit{combinatorial search} techniques. 
For instance, element-wise coordinate descent \cite{wu2019beamforming} sequentially optimizes each RIS element, while phase intercept methods \cite{ren2022linear} evaluate a reduced set of candidate phase shifts to reduce complexity. 

Nevertheless, most of these existing strategies lack strict theoretical performance guarantees and are susceptible to being trapped in local optima. 
Furthermore, a common drawback of these strategies is the difficulty of seamlessly integrating them with well-established continuous optimization theories, often necessitating the development of specialized, problem-specific solvers. 
These observations motivate our work.

Discrete optimization is closely related to the quantization operation, which is inherently a nonlinear mapping from continuous to discrete values. 
For handling nonlinear quantization operation, the Bussgang decomposition \cite{bussgang1952crosscorrelation,demir2020bussgang} provides a powerful tool for statistically linearizing nonlinear operations by decomposing the output of a nonlinear function into a linear signal component and an uncorrelated distortion noise term. 
In particular, this approach has been extensively employed in the context of low-resolution quantizers for massive multiple-input multiple-output (MIMO) systems, where it enables tractable performance analysis and receiver/precoder designs by effectively capturing the quantization effects \cite{mezghani2012capacity,aghdam2019distortion,wan2020generalized,choi2022energy}.

These prior works primarily adopt the Bussgang decomposition as an analysis tool to characterize the distortion introduced by hardware-imposed quantization on physical signals.
In contrast, rather than modeling the quantization effect on an observed signal, we introduce continuous latent variables and establish their statistical connection to the discrete optimization  via the Bussgang decomposition under a reference distribution.
This latent variable formulation replaces the original discrete optimization problem with a distortion-aware continuous surrogate, where well-established continuous optimization techniques become readily applicable.

\subsection{Contributions}
In this paper, we propose a \emph{Bussgang-based latent lifting (BuLL)} method that provides a principled approach to solving discrete optimization problems arising in RIS-aided communication systems.
The main contributions  are summarized as
\begin{itemize}
    \item 
    We develop the BuLL method for discrete optimization that represents discrete variables as quantized outputs of continuous latent variables.
    Using the Bussgang decomposition~\cite{demir2020bussgang}, we derive closed-form Bussgang moments for the considered discrete mapper under a Gaussian reference distribution.
    These moments remain fixed during latent optimization and incorporate reference-averaged discretization effects into a distortion-aware continuous surrogate.
    Our BuLL method extends the role of the Bussgang decomposition from an analysis tool to the construction of a continuous optimization surrogate.

    \item 
    We apply the BuLL method to RIS beam pattern design for uplink channel estimation in a discrete RIS-aided system.
    By lifting the discrete variable to a continuous latent variable, the BuLL framework replaces the discrete-constrained design with a continuous mean square error (MSE) surrogate.
    The surrogate is optimized via the matrix quadratic transform (QT) \cite{shen2018fractional}.

    \item 
    We further extend our BuLL method to RIS-aided beamforming design for downlink sum spectral efficiency (SE) maximization in a multiuser multiple-input single-output (MU-MISO) system.
    We propose a generalized power iteration (GPI)-based phase-shift optimization method for the distortion-aware SE surrogate.
    This eliminates the penalty-parameter tuning required by the existing GPI-based RIS phase-shift optimization in~\cite{oh2026scalable}.

\end{itemize}

\textit{Notation}:
The superscripts $(\cdot)^{\sf T}$, $(\cdot)^{\sf H}$, and $(\cdot)^{-1}$ denote the transpose, Hermitian, and matrix inversion, respectively. 
${\bf{I}}_N$ is the identity matrix of size $N \times N$ and $\bf 0$ is a zero vector with proper dimension.
We let $\bbC^{m \times n}$ be the $m \times n$ dimensional complex space and use ${\rm diag}(\ba)$ for a diagonal matrix with $\ba$ on its diagonal elements.
We use ${\rm Re}\{\cdot\}$ for the element-wise real part operator and ${\rm Im}\{\cdot\}$ for the element-wise imaginary part operator.
$\|\cdot\|$ and $\|\cdot\|_F$ denote the L2 norm of a vector and the Frobenius norm of a matrix, respectively.
We use ${\rm{tr}}(\cdot)$ for trace operator, ${\rm{vec}}(\cdot)$ for vectorization, and $\otimes$ for Kronecker product.
With mean $m$ and variance $\sigma^{2}$, we use $\cC\cN(m,\sigma^{2})$ for a circularly symmetric complex Gaussian distribution.
{We denote a correlation matrix as $\bC_{\bf xy} = \mathbb{E}[\bx\by^{\sf H}]$.}

\section{System Model} \label{sec:sys_model}

We consider a RIS-aided multiuser communication system in which a base station (BS) is equipped with $N$ antennas and serves $K$ single-antenna users.
We assume that direct links between the BS and users are blocked.
The RIS comprises  $M$ reflecting elements characterized by the diagonal phase-shift matrix ${\boldsymbol{\Phi}} = {\rm diag}(\bphi)$ where $\bphi = [\phi_{1}, \phi_{2}, \ldots, \phi_{M}]^{\sf T}$ and $\phi_m = e^{j\theta_m}$ for all $m \in \{1, 2, \ldots, M\}$.
We consider discrete RIS elements with $b$-bit phase resolution. 
To model this discrete nature, we define the set of feasible phase values $\bbF$ with cardinality $L=2^b$, which are uniformly distributed on the unit circle. 
The discrete set is given by
\begin{align}
    \bbF = \left\{ e^{j \frac{2\pi l}{L}} \; \bigg| \; l = 0, 1, \dots, L-1 \right\}.    
\end{align}
For a given $b$-bit, each phase shift is restricted to take one of the discrete values in $\bbF$, i.e., $\phi_m \in \bbF$ for all $m \in \{1, 2, \ldots, M\}$.


In the channel model, we assume a time-division duplexing (TDD) protocol.
Let $\bH_{1} \in \mathbb{C}^{N \times M}$ denote a channel matrix from the RIS to the BS, and let $\bh_{2,k} \in \mathbb{C}^{M}$ denote a channel vector from the $k$-th user to the RIS.
The RIS-user channel matrix is defined as $\bH_{2} = [\bh_{2,1}, \dots, \bh_{2,K}] \in \mathbb{C}^{M \times K}$.
Under the assumption of quasi-static block fading, an effective channel matrix $\bH \in \bbC^{N \times K}$ encompassing RIS phase shifts is given by
\begin{align}
    \label{eq:ch}
    {\bH = \bH_{1} {\boldsymbol{\Phi}} \bH_{2} = [{\bf h}_1, \dots, {\bf h}_K],}
\end{align}
where $\bh_k \in \mathbb{C}^{N}$ represents the effective channel vector of the $k$-th user.

\section{BuLL Discrete Optimization} \label{sec:bussgang}
In this section, we establish a novel discrete optimization framework to address the severe intractability arising from the discrete RIS phase shifters.
The core idea is to lift the target discrete optimization variable into a continuous latent space by exploiting the statistical relationship.
We remark that  such a latent variable does not physically exist in the system.
Specifically, we view each discrete variable as arising from the discretization of an underlying continuous latent variable, which allows the original objective function to be reformulated in terms of the latent variable and discrete mapping. 

\subsection{BuLL Framework \label{subsec:latent_bussgang}}
We begin by reviewing the Bussgang decomposition \cite{bussgang1952crosscorrelation, demir2020bussgang}, which provides a principled framework for statistically linearizing nonlinear operations.
\begin{theorem}[Bussgang decomposition \cite{bussgang1952crosscorrelation, demir2020bussgang}]
    \label{th:Buss_decom}
    Let $\bx$ be a zero-mean complex Gaussian random vector with nonsingular covariance. 
    The output  of a nonlinear  function $\cU(\cdot)$ can be decomposed into a linear signal component of $\bx$ and a distortion noise term $\bd$ as 
    \begin{align}
        \label{eq:general_Bussgang_decom}
        {\bz} = \cU(\bx) = \bB \bx + \bd,
    \end{align}
    where $\bB = \bC_{\bz\bx}\bC_{\bx\bx}^{-1}$ is the Bussgang gain matrix determined in the minimum mean square error (MMSE) sense between $\bz$ and $\bB\bx$, and $\bd$ is the distortion noise satisfying $\bbE[\bd\bx^{\sf H}] = \mathbf{0}$.
\end{theorem}

To understand the conventional use of the Bussgang decomposition, we first present a typical application:

\begin{example}[Low-resolution quantization in massive MIMO]
    \label{ex:mimo_quantization}
    \normalfont
    Theorem~\ref{th:Buss_decom} is widely used to model low-resolution analog-to-digital converters (ADCs) in MIMO receivers \cite{mezghani2012capacity, wan2020generalized, choi2022energy}.
    Consider the received signal $\by_{\rm rx} = \bar{\bH}\bx_{\rm tx} + \bar{\bn} \in \bbC^{N}$, where $\bx_{\rm tx} \sim \cC\cN({\bf 0}, \bR_{\bx\bx})$ and $\bar{\bn} \sim \cC\cN({\bf 0}, \bR_{\bn\bn})$ are independent.
    Since $\by_{\rm rx}$ is a physical signal, its distribution is determined by the signal and noise statistics as $\by_{\rm rx} \sim \cC\cN(\mathbf{0}, \bar{\bH}\bR_{\bx\bx}\bar{\bH}^{\sf H} + \bR_{\bn\bn})$.
    Applying Theorem~\ref{th:Buss_decom} to the ADC output yields
    \begin{align}
        \label{eq:conventional_bussgang}
        \br = \cQ_{\sf adc}(\by_{\rm rx}) =
        \bB_{\sf adc}\by_{\rm rx} + \bd_{\sf adc},
    \end{align}
    where $\bB_{\sf adc}$ is the Bussgang gain matrix and $\bd_{\sf adc}$ is the distortion noise satisfying $\bbE[\bd_{\sf adc}\by_{\rm rx}^{\sf H}] = \mathbf{0}$.
    Here, $\br$ serves as an observation for estimating $\bx_{\rm tx}$ (or $\bar{\bH}$), and the Bussgang decomposition acts as an {\it analytical tool} for characterizing hardware-induced quantization distortion.
\end{example}

Unlike the conventional usage discussed in Example~\ref{ex:mimo_quantization}, we now propose a {BuLL method}; we introduce a continuous latent variable $\breve{\bphi} \in \bbC^{M}$ that does not physically exist in the system and deliberately interpret the discrete phase-shift vector $\bphi \in \bbF^{M}$ as if it were the quantized output of this latent variable.
The detailed steps are presented as follows:

\subsubsection{Introduction of Latent Bussgang Variable} 

    Introducing a continuous latent variable $\breve{\bphi} \in \bbC^M$, we consider the discrete phase-shift vector ${\bphi}$ as an output of a continuous-to-discrete mapper. 
    For the following calibration, $\breve{\bphi}_{\rm ref}$ is a reference draw with i.i.d. $\mathcal{CN}(0,1)$ entries.
    We write $\breve{\bphi}$ for $\breve{\bphi}_{\rm ref}$ for brevity, and all expectations refer to this reference law.
    Accordingly, Theorem~\ref{th:Buss_decom} gives
    \begin{align}
        \label{eq:Bussgang_decom}
        {\bphi}_{\rm ref} = \cQ(\breve{\bphi}) = \bA \breve{\bphi} + \bq,
    \end{align}
    where $\cQ(\cdot)$ denotes a nonlinear continuous-to-discrete mapper that maps each input entry to a feasible value in $\bbF$, $\breve{\bphi} = [\breve{{\phi}}_1, \ldots, \breve{{\phi}}_M]^{\sf T} \in \bbC^{M}$ denotes the continuous latent variable, $\bA \in \mathbb{C}^{M \times M}$ is the corresponding Bussgang gain matrix, and $\bq \in \mathbb{C}^{M}$ is the distortion noise satisfying $\bbE[\bq\breve{\bphi}^{\sf H}] = \mathbf{0}$.
    In addition, $\bbE[\bq] = {\bf 0}$ holds because $\bbE[\bphi_{\rm ref}] = \bbE[\breve{\bphi}] = {\bf 0}$.

    The Bussgang decomposition essentially models the nonlinear output as a linear transform of the input plus an error term. 
    To ensure that the linear model $\bA\breve{\bphi}$ linearly expresses ${\bphi}_{\rm ref}$ in the MMSE sense, $\bq = {\bphi}_{\rm ref} - \bA\breve{\bphi}$ must be uncorrelated to the input $\breve{\bphi}$.
    Thus, imposing $\bbE[\bq\breve{\bphi}^{\sf H}] = \mathbf{0}$ yields:
    \begin{align}
        \bbE[({\bphi}_{\rm ref} - \bA\breve{\bphi})\breve{\bphi}^{\sf H}] = \bC_{{\bphi}\breve{\bphi}} - \bA\bC_{\breve{\bphi}\breve{\bphi}} = \mathbf{0},
    \end{align}
    which implies $\bA = \bC_{{\bphi}\breve{\bphi}}\bC_{\breve{\bphi}\breve{\bphi}}^{-1}$. 
    The distortion noise covariance matrix is given by
    \begin{align}
        \bC_{\bq\bq} = \bbE[(\bphi_{\rm ref} - \bA\breve{\bphi})
        (\bphi_{\rm ref} - \bA\breve{\bphi})^{\sf H}].
    \end{align}
    Using the uncorrelation $\bbE[\bq\breve{\bphi}^{\sf H}] = \mathbf{0}$, 
    this simplifies to
    \begin{align}
        \label{eq:Bussgang_Cqq}
        \bC_{\bq\bq} = \bC_{\bphi\bphi} - \bA\bC_{\breve{\bphi}\breve{\bphi}}
        \bA^{\sf H}.
    \end{align}

\begin{table*}[t]
    \centering
    \caption{Comparison between Conventional Bussgang and Proposed 
   BuLL Approaches}
    \vspace{-0.6em}
    \label{tab:bussgang_contrast}
    \renewcommand{\arraystretch}{1}
    \begin{tabular}{@{} l | p{6.0cm} | p{6.0cm} @{}}
        \toprule
        \textbf{Aspect}
        & \textbf{Conventional Bussgang Approach} 
        (Example~\ref{ex:mimo_quantization})
        & \textbf{Proposed BuLL Approach} 
        \\
        \midrule
        Optimization target
        & $\bx_{\rm tx}$ or $\bar{\bH}$ (not a quantized output)
        & $\bphi$ discrete phase shifts (quantized output)
        \\[3pt]
        Physical existence of quantizer input
        & $\by_{\rm rx}$ physically exists as a received signal
        & $\breve{\bphi}$ is fictitious; 
          no physical counterpart 
        \\[3pt]
        Quantizer input distribution
        & Determined by signal, channel, and noise statistics
        & Deliberately chosen by designer 
          (e.g., $\mathcal{CN}(\mathbf{0},\mathbf{I}_M)$) 
        \\[3pt]
        Role of Bussgang decomposition
        & \emph{Analytical tool}: characterizes hardware-imposed 
          quantization distortion
        & \emph{Optimization tool}: transforms discrete constraint 
          into continuous optimization via stochastic relaxation
        \\[3pt]
        Bussgang gain \& distortion covariance
        & Reflects actual ADC quantization level
        & Controlled by the choice of latent distribution 
        \\
        \bottomrule
    \end{tabular}
\end{table*}

\subsubsection{Discrete Mapper Design and Latent Distribution Selection}

    To compute the Bussgang gain matrix $\bA$ and the distortion noise covariance matrix $\bC_{\bq\bq}$, we need to specify two design choices: $(i)$ the distribution of the latent variable $\breve{\bphi}$ and $(ii)$ the form of the continuous-to-discrete mapper $\cQ(\cdot)$.
    Unlike the conventional setting in Example~\ref{ex:mimo_quantization}, where the input distribution is naturally determined by the channel and noise statistics, no physical model governs $\breve{\bphi}$ in our formulation. 
    Both choices must therefore be made jointly by the designer so that the resulting mapping $\cQ(\breve{\bphi})$ produces a valid discrete phase-shift vector $\bphi \in \bbF^{M}$ while admitting a tractable closed-form expression for $\bA$ and $\bC_{\bq\bq}$.

    For the reference latent distribution, we assume that $\breve{\bphi}$ follows the circularly symmetric complex Gaussian distribution with i.i.d. entries, i.e., $\breve{\bphi} \sim \mathcal{CN}(\mathbf{0}, \bI_M)$. 
    This choice is motivated by the following facts:
    $(i)$ Theorem~\ref{th:Buss_decom} requires a Gaussian input,
    $(ii)$ the zero mean and unit variance match the first and second moments of uniformly distributed reference phase shifts,
    specifically, $\bbE[\phi_m] = 0$ and $\bbE[|\phi_m|^2] = 1$ when $\phi_m$ is uniform on $\bbF$, and 
    $(iii)$ the circularly symmetric complex Gaussian distribution provides a uniform phase distribution.
    Consequently, the Bussgang gain and distortion noise covariance matrices admit the simple closed-form expressions \cite{demir2020bussgang}:
    \begin{align}
        \label{eq:Bussgang_gain_C_qq}
        \bA = \beta \bI_M, \;\; \bC_{\bq\bq} = (1 - \beta^2)\bI_M,
    \end{align}
    where the scalar gain $\beta$ depends only on the resolution $b$.

    For the discrete mapper, we adopt an element-wise phase quantizer that retains only the phase component of each entry:
    \begin{align}
        \label{eq:phase_quantizer}
        \cQ(\breve{\phi}_m) = \cF\!\left(\angle \breve{\phi}_m \right) 
        = \phi_m \in \bbF,
    \end{align}
    where $\cF: [-\pi, \pi) \to \bbF$ is a uniform phase quantizer that maps an input angle to the element of $\bbF$ whose phase is closest to it and $\angle$ denotes the angle  of a complex number.
    Since the target discrete variable $\bphi$ lies on the unit circle, the amplitude of $\breve{\phi}_m$ is irrelevant and is therefore discarded.

\subsubsection{Computation of Latent Bussgang Gain}

Now, we need to compute the gain $\beta$ and the distortion noise covariance $\bC_{\bf qq}$ based on the selection of the  discrete mapper and latent distribution.
The following proposition presents the result for a general $b$-bit phase resolution.

\begin{proposition}[Latent Bussgang gain]
    \label{prop:Bussgang_gain}
    Under the assumption that the continuous latent variable follows  $\breve{\bphi} \sim \mathcal{CN}(\mathbf{0}, \mathbf{I}_M)$, the Bussgang gain $\beta$ for a $b$-bit phase resolution is derived as
    \begin{align}
        \label{eq:Bussgang_gain}
        \beta = \frac{L\sqrt{\pi}}{2\pi} \sin\left(\frac{\pi}{L}\right),
    \end{align}
     where $L=2^b$.
    \begin{proof}
        Please refer to Appendix~\ref{appen:Buss_gain}.
    \end{proof}

\end{proposition}

\begin{remark}[On the choice of reference distribution]\normalfont
We adopt the Gaussian reference, allowing direct application of Theorem~\ref{th:Buss_decom}.
With the element-wise discrete mapper, this choice yields the closed-form gain in Proposition~\ref{prop:Bussgang_gain}
and the distortion covariance $\mathbf C_{\bq\bq}=(1-\beta^2)\mathbf I_M$.
Its uniform phase distribution is also consistent with the uniformly spaced alphabet $\bbF$.
These reference moments remain fixed during optimization.
Non-Gaussian reference distributions with zero mean
and second moments also admit a best linear projection
with an uncorrelated residual.
The corresponding gain and distortion moments depend on the
chosen reference distribution and may require numerical evaluation.
A systematic study of non-Gaussian
reference distributions is left for future work.
\end{remark}

Our BuLL method differs fundamentally from the conventional usage of the Bussgang decomposition, and the key differences are summarized in Table~\ref{tab:bussgang_contrast}. 
In the conventional setting of Example~\ref{ex:mimo_quantization}, the quantized output $\br$ is merely an observation, and the true optimization target is $\bx_{\rm tx}$ or $\bar{\bH}$. 
In contrast, in our approach, the quantized output $\bphi$ itself is the optimization variable, and the Bussgang decomposition is applied to recast it in terms of the continuous latent variable $\breve{\bphi}$. 
Furthermore, the quantizer input $\by_{\rm rx}$ in Example~\ref{ex:mimo_quantization} physically exists as a received signal in the hardware, whereas $\breve{\bphi}$ is a fictitious variable with no physical counterpart and is introduced solely to enable continuous optimization. 
As a result, the role of the Bussgang decomposition shifts from an analytical tool to an optimization tool that transforms discrete constraints into a tractable continuous problem, with the Bussgang gain controlled by the designer's choice of the latent distribution rather than reflecting an actual physical quantization system.

\subsection{BuLL for Discrete Optimization}
\begin{figure*}[t]
    \centering
    \includegraphics[width=0.85\textwidth]{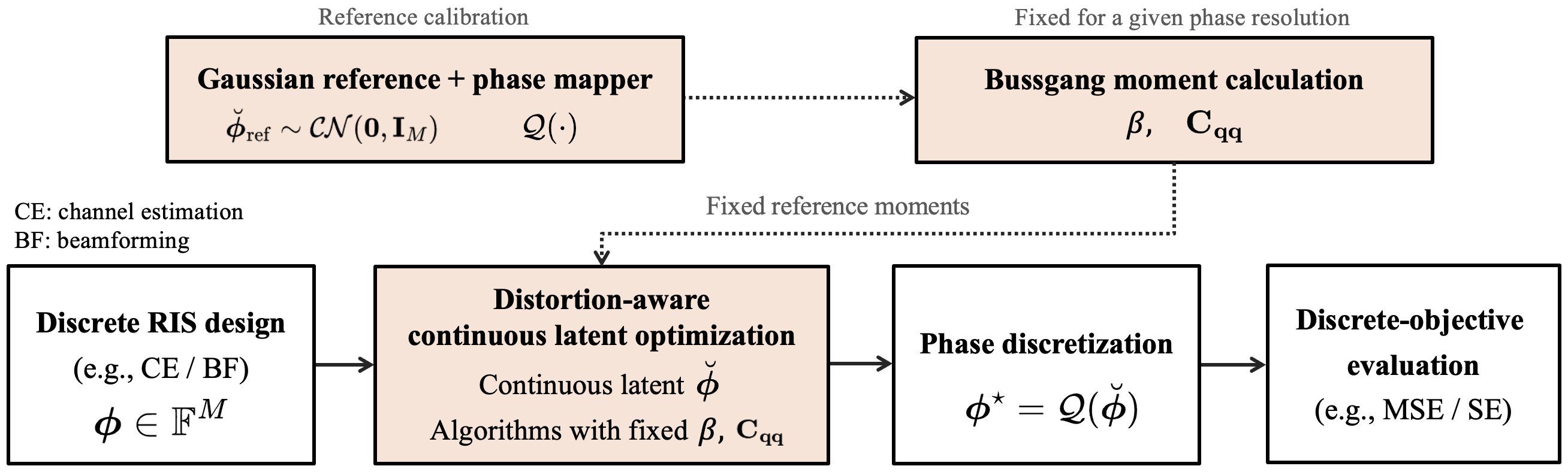}
    \vspace{-0.25cm}
    \caption{A block diagram of BuLL discrete phase-shift optimization for RIS-aided communications.}
    \label{fig:BuLL_frame}
\end{figure*}

As illustrated in Fig.~\ref{fig:BuLL_frame}, the Gaussian reference and the discrete mapper first determine the Bussgang gain $\beta$ and distortion covariance $\mathbf C_{\bq\bq}=(1-\beta^2)\mathbf I_M$, which remain fixed for a given phase resolution.
Using the BuLL framework, we construct a continuous surrogate
over the latent variable $\breve{\bphi}$ instead of directly
optimizing the discrete variable ${\bphi}$.
The latent variable is optimized deterministically under
a latent-energy constraint and then quantized into a feasible
discrete candidate
${\bphi^{\star}} = \cQ(\breve{\bphi}^{\star})$.
The final block in Fig.~\ref{fig:BuLL_frame} evaluates this candidate
using the actual MSE or SE.
The application-specific algorithms retain the best evaluated
feasible design.

Our BuLL and conventional relaxation-then-quantization methods~\cite{wu2019beamforming_icassp, so2007approximating, di2020hybrid} share a broad three-step structure: $(i)$ introduce continuous variables, $(ii)$ optimize the continuous objective, and $(iii)$ map the resulting design to discrete values.
Conventional methods, such as semidefinite relaxation~\cite{gao2021robust} and manifold optimization~\cite{manifold_ref, guo2020weighted, pan2022overview}, when used in this paradigm, optimize a continuous objective without explicitly accounting for the subsequent discretization error, which can lead to performance loss at low phase resolutions.
In contrast, the BuLL method uses a prescribed reference distribution to characterize the statistical relationship between a continuous latent variable and its quantized output.
Under the Gaussian reference, the Bussgang decomposition gives $\bphi_{\rm ref} =\beta\breve{\bphi} + \bq$, where $\beta$ and $\mathbf C_{\bq\bq}$ characterize the linear component and average residual distortion of the mapping.
These fixed moments are incorporated into the effective signal model that drives the continuous objective, allowing reference-averaged discretization effects to be considered during latent optimization before the final phase quantization.
The reference law is used for calibration, and the latent variable is subsequently optimized deterministically.
This construction provides a distortion-aware continuous surrogate for discrete RIS design.

In the following sections, we apply our BuLL method to develop
tractable channel estimation and beamforming algorithms in
discrete-phase RIS-aided communication systems.

\section{RIS Optimization for Channel Estimation } \label{sec:CE}
In this section, we consider the uplink channel estimation problem for a discrete-phase RIS-aided multiuser single-input multiple-output (MU-SIMO) system.
Our primary goal is to accurately estimate cascaded channels from the users to the BS via the RIS with discrete phase shifts.
To this end, we design the RIS beam pattern to enhance channel estimation accuracy. 
Leveraging our BuLL method in Section~\ref{sec:bussgang}, we reformulate the problem over continuous latent variables and solve it efficiently via a QT approach.

\subsection{Problem Formulation under BuLL Framework}
In the channel estimation process, we assume the transmission is divided into $T$ blocks where $T\geq M$. In each block with $n \in \{1, 2, \dots, T\}$, we consider an orthonormal pilot sequence with $T_s\geq K$ time slots for user $k$ denoted as
\begin{align}
    \bt_{k}[n] = \big[t_{k,1}[n], t_{k,2}[n], \ldots, t_{k,T_s}[n]\big]^{\sf T}\in \bbC^{T_s},\;\forall k.
\end{align}
With these pilots, the received uplink signal at the BS in the $n$-th block is given by
\begin{align}
    \bY^{\sf UL}[n] &= \sum_{k=1}^{K}\sqrt{\rho}\,\bH^{\sf r}_{k}\bphi[n]\, \bt_{k}^{\sf H}[n] + \bN[n] \in \bbC^{N \times T_s},
\end{align}
where $\bphi[n]$ is the discrete RIS phase-shift vector applied during the $n$-th block, $\bH^{\sf r}_k$ is the cascaded channel of user $k$ defined as $\bH^{\sf r}_{k} = \bH_{1} {\rm diag}(\bh_{2,k}) \in \bbC^{N \times M}$, and $\rho$ is the common uplink training power across all users.
To remove inter-user interference, we correlate the received signal with the orthonormal pilot sequence of user $k$ as
\begin{align}
    \by^{\sf UL}_{k}[n] &= \bY^{\sf UL}[n]\bt_{k}[n] 
    = \sqrt{\rho} \bH^{\sf r}_{k} \bphi[n] + {\bn}_{k}[n]
    \nonumber \\
    &= \sqrt{\rho}\left(\bphi^{\sf T}[n] \otimes \bI_N \right)\bc_{k} + {\bn}_{k}[n],
    \label{eq:Y_OS}
\end{align}
where ${\bn}_{k}[n] = \bN[n] \bt_k[n]$ and $\bc_k = {\rm vec}(\bH^{\sf r}_k) \in \bbC^{MN}$.
We stack the uplink signals for $n=1,2,\ldots, T$ as
\begin{align}
    \nonumber
    \widebar{\by}_{k} &=
    \begin{bmatrix}
        {\by}_{k}^{\sf UL}[1] 
        \\
        \vdots 
        \\
        {\by}_{k}^{\sf UL}[T] 
    \end{bmatrix}
   = \sqrt{\rho} 
    \begin{bmatrix}
        \left(\bphi^{\sf T}[1] \otimes \bI_N \right)  
        \\
        \vdots  
        \\
        \left(\bphi^{\sf T}[T] \otimes \bI_N \right)  
    \end{bmatrix}
    \bc_{k} + 
    \begin{bmatrix}
        {\bn}_{k}[1]  
        \\
        \vdots  
        \\
        {\bn}_{k}[T]  
    \end{bmatrix}
    \\
    &= \bZ \bc_{k} + \bar{\bn}_{k} \in \bbC^{N T},
\end{align}
where
\begin{align}
    \boldsymbol{\Psi} &=
        [\bphi[1]
        ,\dots,  
        \bphi[T]]^{\sf T} \in \bbC^{T \times M},
    \\
    \bZ &= \sqrt{\rho} \left(\boldsymbol{\Psi} \otimes \bI_N \right)\in \bbC^{NT \times NM}.
\end{align}
To estimate $\bc_k$ from the stacked observations, we employ a linear minimum mean square error (LMMSE) estimator \cite{pan2022overview, kundu2021channel}, which is given by
\begin{align}
    \label{eq:LMMSE}
    \bW^{\rm LMMSE}_{k} &= \bC_{\bc_k \bar{\by}_k}\bC_{\bar{\by}_k \bar{\by}_k}^{-1}
    = \bC_{k} \bZ^{\sf H} \left( \bZ\bC_{k}\bZ^{\sf H} + \sigma^2 \bI_{NT} \right)^{-1},
\end{align}
where $\bC_{k} = \bbE[\bc_k \bc_k^{\sf H}]$ is the cascaded channel covariance matrix.
Using \eqref{eq:LMMSE}, the estimated channel is given by
\begin{align}
    \bW^{\rm LMMSE}_{k}\, \widebar{\by}_{k} = \hat{\bc}_{k} \in \bbC^{NM}.
\end{align}
We define the channel estimation error vector as $\be_k = \bc_k - \hat{\bc}_k$.
Then, the LMMSE estimation error covariance  is
\begin{align}
    \label{eq:err_cov}
    \bR^{\sf e}_{k} = \bbE\left[\be_k\be^{\sf H}_k\right] =  \left(\bC_k^{-1} + \frac{\rho}{\sigma^2} {\boldsymbol{\boldsymbol{\Psi}}}^{\sf H}{\boldsymbol{\boldsymbol{\Psi}}} \otimes \bI_N \right)^{-1}.
\end{align}
Subsequently, the $k$th user's MSE is defined as
\begin{align}
    \label{eq:MSE}
    {\rm MSE}_{k} = {\rm tr}\big(\bR^{\sf e}_k\big).
\end{align}
Thus, the RIS pattern design problem can be formulated as
\begin{align} 
    \label{eq:main_problem}
    \mathop{{\text{minimize}}}_{{\boldsymbol{\Psi}}}& \;\; \sum_{k=1}^{K}{\rm MSE}_k 
    \\
    \label{eq:unitmodulus_constraint}
    \mathop{{\text{subject to}}}& \;\; \left[{\boldsymbol{\Psi}}\right]_{t,m} \in \bbF,\;\forall t,m.
\end{align}
The discrete-phase constraint in \eqref{eq:unitmodulus_constraint} renders \eqref{eq:main_problem} an MINLP, which is generally NP-hard \cite{so2007approximating,burer2012non,wu2019beamforming}.
For a $b$-bit discrete-phase RIS, the discrete Fourier transform (DFT) matrix is optimal when $M \in \{2^c \mid c=1,\ldots,b\}$ \cite{you2020channel}.
Under 1-bit phase quantization, a Hadamard matrix, when available, is also optimal \cite{you2020channel}.

To accommodate general RIS sizes under the discrete-phase constraint, we apply our BuLL method.
Using the fixed reference moments in \eqref{eq:Bussgang_decom}, we  introduce the working model $\boldsymbol{\Psi}=\beta\widebreve{\boldsymbol{\Psi}}+\mathbf{Q}$, where 
\begin{align}
    \widebreve{\boldsymbol{\Psi}} &\!=\! 
   \left[\boldsymbol{\breve\phi}[1],\cdots, \boldsymbol{\breve\phi}[T]\right]^{\sf T}\in \mathbb{C}^{T \times M}
\end{align}
is the deterministic latent matrix.
The vectorized $\mathrm{vec}(\bQ)$ has zero mean and the fixed reference covariance $\bC_{\bq\bq}\otimes\bI_T$.
Under this working model, we consider the expected error covariance matrix $\bbE_{\bQ} [\bR_k^{\sf e}]$. 
However, obtaining a closed-form expression for this expectation is intractable due to the matrix inversion involved. 
To obtain a tractable surrogate, we apply Jensen's inequality, exploiting the matrix convexity of the inverse function $\bbE[\bA^{-1}] \succeq \bbE[\bA]^{-1}$ \cite{bhatia2013matrix}:
\begin{align}
    \nonumber
    &\bbE_{\bQ}[\bR_k^{\sf e}(\boldsymbol{\Psi})] \succeq  \bbE_{\bQ}\!\left[\bC_k^{-1} \!+\! \frac{\rho}{\sigma^2} \boldsymbol{\Psi}^{\sf H}\boldsymbol{\Psi} \otimes \bI_N \right]^{-1}
    \\
    \label{eq:LB_Cov}
    & \qquad = \left(\tilde{\mathbf{C}}_k^{-1} + \frac{\rho}{\sigma^2} (\beta \widebreve{\boldsymbol{\Psi}} \otimes \bI_N)^{\sf H}(\beta \widebreve{\boldsymbol{\Psi}} \otimes \bI_N) \right)^{-1} = \Tilde{\bR}_k^{\sf e},
\end{align}
where $\tilde{\mathbf{C}}_k = \left(\mathbf{C}_k^{-1} + \frac{\rho}{\sigma^2} (T \mathbf{C}_{\bq\bq} \otimes \mathbf{I}_N)\right)^{-1}$.
By applying the Woodbury matrix identity \cite{woodbury1950inverting}, $\Tilde{\bR}_k^{\sf e}$ can be  expressed as
\begin{align}
    \label{eq:R_e_SWM}
    \Tilde{\bR}_k^{\sf e} = \tilde{\mathbf{C}}_k - \tilde{\mathbf{C}}_k \mathbf{X}^{\sf H} (\mathbf{X} \tilde{\mathbf{C}}_k \mathbf{X}^{\sf H} + \mathbf{W})^{-1} \mathbf{X} \tilde{\mathbf{C}}_k,
\end{align}
where $\bX = \beta \widebreve{\boldsymbol{\Psi}} \otimes \bI_N$ and $\bW = \frac{\sigma^2}{\rho}\bI_{NT}$.


We minimize $\sum_{k=1}^{K}{\rm tr}(\tilde{\bR}_k^{\sf e})$
as a design surrogate for the expected sum MSE.
To define the continuous feasible set, note that every discrete RIS pattern has unit-modulus entries and thus satisfies $\|\boldsymbol{\Psi}\|_F^2=TM$.
We relax the entry-wise discrete restrictions to a convex energy ball for the latent matrix, using $TM$ as the total-energy budget.
Dropping the constant term then yields the following surrogate maximization problem:
\begin{align}
    \label{eq:main_problem2}
    \mathop{\text{maximize}}_{\widebreve{\boldsymbol{\Psi}}}\quad
    &\cL(\widebreve{\boldsymbol{\Psi}})
      \triangleq\sum_{k=1}^{K}\cL_k(\widebreve{\boldsymbol{\Psi}}),\\
    \label{eq:CE_energy_constraint}
    \text{subject to}\quad
    &\|\widebreve{\boldsymbol{\Psi}}\|_F^2\leq TM,
\end{align}
where $\cL_k(\widebreve{\boldsymbol{\Psi}})
    =\operatorname{tr}\!\left(
      \tilde{\bC}_k\bX^{\sf H}
      (\bX\tilde{\bC}_k\bX^{\sf H}+\bW)^{-1}
      \bX\tilde{\bC}_k\right).$
The latent-energy constraint in \eqref{eq:CE_energy_constraint} contains all discrete feasible patterns.
The budget $TM$ also matches the average total energy of the Gaussian reference matrix, i.e., $\bbE\left[\|\widebreve{\boldsymbol{\Psi}}_{\rm ref}\|_F^2\right]=TM$.

\subsection{Quadratic Transform} \label{subsec:CE_FP}



The continuous surrogate in \eqref{eq:main_problem2} is nonconvex because it contains matrix ratios.
To resolve this, we use the quadratic transform \cite{shen2018fractional}  under \eqref{eq:CE_energy_constraint}.

\begin{theorem}
    \label{th:matix_QT}
    (Matrix Quadratic Transform~\cite{shen2018fractional,shen2018fractional2,shen2020enhanced}):
    Let $\CMcal{X}$ be a nonempty constraint set.
    For $j=1,\ldots,J$, consider functions $\bA_j(\bx) \in \bbC^{a_1 \times a_2}$ and positive-definite matrix-valued functions $\bB_j(\bx) \in \bbH^{a_1 \times a_1}$.
    Let $F_j: \bbH^{a_2 \times a_2} \mapsto \bbR$ satisfy $F_j(\bC) \geq F_j(\bC^{'})$ whenever $\bC \succeq \bC^{'}$.
    The sum-of-functions-of-matrix-ratio problem
    \begin{align}
        \mathop{{\textnormal{maximize}}}_{\bx}& \;\; \sum_{j=1}^{J} F_j \left(\bA_j^{\sf H}(\bx)\bB_j^{-1}(\bx)\bA_j(\bx) \right)
        \\
        \mathop{{\textnormal{subject to}}}& \;\; \bx \in \CMcal{X}
    \end{align}
    is equivalent to 
\begin{align}
        \label{eq:main_FP_QT}
        \mathop{{\textnormal{maximize}}}_{\bx, \underline{\bY}}& \;\; \sum_{j=1}^{J} F_j \left(\bA_j^{\sf H}(\bx)\bY_j \!+\bY_j^{\sf H}\bA_j(\bx) \!-\! \bY_j^{\sf H}\bB_j(\bx)\bY_j \right)
        \\
        \mathop{{\textnormal{subject to}}}& \;\; \bx \in \CMcal{X},\;\; \bY_j \in \bbC^{a_1 \times a_2},
    \end{align}
where $\bbH^{m \times m}$ denotes the set of $m \times m$ Hermitian matrices, $\bY_j$ is an auxiliary variable introduced for each matrix ratio term $\bA_j^{\sf H}(\bx)\bB_j^{-1}(\bx)\bA_j(\bx)$, and $\underline{\bY} = \left[\bY_1, \bY_2, \ldots, \bY_J\right]$.
\end{theorem}
Applying Theorem~\ref{th:matix_QT} to our problem, we obtain the following equivalent reformulation.
\begin{proposition}
    Employing the matrix quadratic transform in Theorem~\ref{th:matix_QT}, the problem in \eqref{eq:main_problem2} is transformed to 
    \begin{align}
        \mathop{{\textnormal{maximize}}}_{\widebreve{\boldsymbol{\Psi}}, \underline{\bY}}& \;\; f(\widebreve{\boldsymbol{\Psi}}, \underline{\bY})
        \\
        \mathop{{\textnormal{subject to}}}& \;\; \bY_k \in \bbC^{NT \times NM},
        \\
        &\|\widebreve{\boldsymbol{\Psi}}\|_F^2\leq TM,
    \end{align}
    where $f(\widebreve{\boldsymbol{\Psi}}, \underline{\bY}) \!=\! \sum_{k=1}^{K} f_k$ 
    and $f_k = {\rm tr}\left(2{\rm Re}\{\Tilde{\bC}_k \bX^{\sf H}\bY_k \} \right) \!-\! {\rm tr}\left(\bY_k^{\sf H}\left(\bX \Tilde{\bC}_k \bX^{\sf H} + \bW \right)\bY_k \right)\!.$
    \begin{proof}
        The reformulation is obtained by treating $\bX \Tilde{\bC}_k$ and $\bX\Tilde{\bC}_k\bX^{\sf H} + \bW $ as $\bA_j$ and $\bB_j$ in Theorem~\ref{th:matix_QT}, respectively.
        The matrix $\bX\Tilde{\bC}_k\bX^{\sf H} + \bW$ is positive definite for all $k$.
        The trace function satisfies the monotonicity condition in Theorem~\ref{th:matix_QT} because ${\rm tr}(\bC) \geq {\rm tr}(\bC^{'})$ whenever $\bC \succeq \bC^{'}$.
    \end{proof}
\end{proposition}

We solve this reformulated problem by alternately optimizing the auxiliary variables $\underline{\mathbf{Y}}$ and the latent matrix $\widebreve{\boldsymbol{\Psi}}$ subject to \eqref{eq:CE_energy_constraint}.
For fixed $\widebreve{\boldsymbol{\Psi}}$, $f$ is concave in the auxiliary variables.
Setting $\partial f / \partial \mathbf{Y}_k= \mathbf{0}$ gives
\begin{align}
    \label{eq:Y_k_star}
    \bY_k^{\star} = \left(\bX\Tilde{\bC}_k\bX^{\sf H} + \bW \right)^{-1}\bX\Tilde{\bC}_k.
\end{align}
For fixed $\underline{\bY}$, $f$ is concave in the latent matrix.
The energy constraint in \eqref{eq:CE_energy_constraint} is convex.
Hence, the constrained latent subproblem can be solved globally.
Let us define
\begin{align}
    \label{eq:R_k} \bR_k=\bY_k\bY_k^{\sf H},\;\;\bS_k=\bY_k\tilde{\bC}_k,\;\;\pmb{\psi}={\rm vec}(\widebreve{\boldsymbol{\Psi}}).
\end{align}
We then introduce the fixed selection matrix $\bL_N$ satisfying $\bL_N\pmb{\psi}={\rm vec}(\widebreve{\boldsymbol{\Psi}}\otimes\bI_N)$, which gives $\bL_N^{\sf H}\bL_N= N \bI_{TM}$.
Subsequently, the variable-dependent part of $f$ is defined as
\begin{align}
    \label{eq:PSI_new_main}
    \tilde{f}(\widebreve{\boldsymbol{\Psi}}, \underline{\bY}) = 2\beta\operatorname{Re}\{\pmb{\psi}^{\sf H}\ba_{\psi}\}-\beta^2\pmb{\psi}^{\sf H}\bH_{\psi}\pmb{\psi},
\end{align}
where 
\begin{align}
    \label{eq:CE_QT_coefficients}
    \bH_{\psi}\!=\!\sum_{k=1}^{K}\bL_N^{\sf H}(\tilde{\bC}_k^{\sf T}\otimes\bR_k)\bL_N, \;\; \ba_{\psi}\!=\!\sum_{k=1}^{K}\bL_N^{\sf H}{\rm vec}(\bS_k).
\end{align}
With \eqref{eq:PSI_new_main}, we now rewrite the optimization problem as
\begin{align}
    \mathop{\text{maximize}}_{\pmb{\psi}}\;\;
    &2\beta\operatorname{Re}\{\pmb{\psi}^{\sf H}\ba_{\psi}\} - \beta^2\pmb{\psi}^{\sf H}\bH_{\psi}\pmb{\psi}
    \\
    \text{subject to}\;\;
    &\|\pmb{\psi}\|^2  \leq TM,
\end{align}
where the objective is concave and the feasible set is convex.
Since $\pmb{\psi}={\bf 0}$ is strictly feasible, Slater's condition holds, and the following Karush–Kuhn–Tucker conditions are necessary and sufficient for global optimality of the latent subproblem:
\begin{align}
    \label{eq:CE_QT_KKT}
    &(\beta^2\bH_{\psi}+\lambda\bI_{TM})\pmb{\psi}^{\star}
       =\beta\ba_{\psi},\;\;\; \lambda\geq0,\\
    &\|\pmb{\psi}^{\star}\|^2\leq TM,\;\;\;
    \lambda(\|\pmb{\psi}^{\star}\|^2-TM)=0.\nonumber
\end{align}
Let $\pmb{\psi}_0=\beta^{-1}\bH_{\psi}^{\dagger}\ba_{\psi}$ be the minimum-norm unconstrained maximizer, where $\dagger$ denotes the Moore--Penrose inverse.
If $\pmb{\psi}_0$ is feasible, choose $\lambda=0$ and $\pmb{\psi}^{\star}=\pmb{\psi}_0$.
Otherwise, the constraint is active and the optimal solution is 
\begin{align}
    \label{eq:Phi_star}
    \pmb{\psi}^{\star} =(\beta^2\bH_{\psi}+\lambda\bI_{TM})^{-1}\beta\ba_{\psi}.
\end{align}
Using the eigendecomposition
$\bH_{\psi}=\bU\operatorname{diag}(d_1,\ldots,d_{TM})\bU^{\sf H}$
and defining
$\bz=\bU^{\sf H}\ba_{\psi}=[z_1,\ldots,z_{TM}]^{\sf T}$,
the unique positive multiplier satisfies
\begin{align}
    \label{eq:CE_lambda_root}
    \sum_{i=1}^{TM}
    \frac{\beta^2|z_i|^2}{(\beta^2d_i+\lambda)^2}
    =TM.
\end{align}
For general eigenvalues $d_i$ in \eqref{eq:CE_lambda_root}, this scalar equation
does not admit a simple closed-form solution.
Since its left-hand side is continuous and strictly decreasing
in $\lambda>0$, we solve \eqref{eq:CE_lambda_root} numerically
by bisection to a prescribed tolerance, using the bracket
$[0,\beta\|\ba_{\psi}\|/\sqrt{TM}]$, where the left-hand side of \eqref{eq:CE_lambda_root} is bounded above by $\beta^2\|\ba_{\psi}\|^2/\lambda^2$ for $\lambda>0$.
We summarize these steps in Algorithm~\ref{alg:algorithm_QT}, where quantization is used only to evaluate discrete candidates without affecting the continuous latent updates, and ${\rm MSE}_{\Sigma}(\boldsymbol{\Psi})$ denotes the sum MSE evaluated for the RIS pattern $\boldsymbol{\Psi}$.
The retained pattern is replaced only when a candidate achieves a lower actual MSE.


\begin{algorithm}[t]
    \caption{BuLL QT-based RIS beam pattern Optimization}
    \label{alg:algorithm_QT}
    Initialize nonzero $\widebreve{\boldsymbol{\Psi}}^{(0)}$ satisfying \eqref{eq:CE_energy_constraint}.\;
    Set $\boldsymbol{\Psi}_{\rm best}\!=\!\cQ(\widebreve{\boldsymbol{\Psi}}^{(0)})$, $o_{\rm best}\!=\!{\rm MSE}_{\Sigma}(\boldsymbol{\Psi}_{\rm best})$, $L_0\!=\!\cL(\widebreve{\boldsymbol{\Psi}}^{(0)})$, $t=1$, and $\Delta_0=\infty$.\;
    \While{$\Delta_{t-1}>\varepsilon_1$ \textbf{\&} $t\leq t_{1,\max}$}{
        Compute $\bY_k$ using \eqref{eq:Y_k_star} at $\widebreve{\boldsymbol{\Psi}}^{(t-1)}$.\;
        Form $\bR_k,\bS_k$ in $\eqref{eq:R_k}$, and then $\bH_{\psi},\ba_{\psi}$ in \eqref{eq:CE_QT_coefficients}.\;
        Set $\pmb{\psi}_0=\beta^{-1}\bH_{\psi}^{\dagger}\ba_{\psi}$.\;
        \eIf{$\|\pmb{\psi}_0\|^2\leq TM$}{
            Set $\lambda_t=0$ and $\pmb{\psi}^{(t)}=\pmb{\psi}_0$.\;
        }{
            Find $\lambda_t>0$ from \eqref{eq:CE_lambda_root} by bisection.\;
            Compute $\pmb{\psi}^{(t)}$ using \eqref{eq:Phi_star} with $\lambda=\lambda_t$.\;
        }
        Set $\widebreve{\boldsymbol{\Psi}}^{(t)}={\rm unvec}(\pmb{\psi}^{(t)})$.\;
        Set $\boldsymbol{\Psi}_{\rm d}=\cQ(\widebreve{\boldsymbol{\Psi}}^{(t)})$.\;
        \If{${\rm MSE}_{\Sigma}(\boldsymbol{\Psi}_{\rm d})<o_{\rm best}$}{
            $\boldsymbol{\Psi}_{\rm best}\leftarrow\boldsymbol{\Psi}_{\rm d}$ and $o_{\rm best}\leftarrow{\rm MSE}_{\Sigma}(\boldsymbol{\Psi}_{\rm d})$.\;
        }
        Set $L_t=\cL(\widebreve{\boldsymbol{\Psi}}^{(t)})$ and $\Delta_t=|L_t-L_{t-1}|/|L_{t-1}|$.\;
        $t\leftarrow t+1$.\;
    }
    \Return{$\boldsymbol{\Psi}^{\star}=\boldsymbol{\Psi}_{\rm best}$}.
\end{algorithm}

\subsection{Complexity and Convergence Analysis} \label{subsec:CE_complexity}

For dense matrices with $T\geq M$, Algorithm~\ref{alg:algorithm_QT}
has $\CMcal{O}\!\left(KN^3T^3+KN^2T^2M^2+(TM)^3+J_{\lambda}TM\right)$ per-iteration complexity.
The last two terms account for eigendecomposition of $\bH_{\psi}$ and $J_{\lambda}$ scalar bisection evaluations, respectively.
After coefficient formation, an optional Fast-QT implementation~\cite{shen2024accelerating} can reduce the cost per latent-update step to $\CMcal{O}((TM)^2)$.

Assume fixed $\tilde{\bC}_k\succ\mathbf{0}$, $\beta,\rho,\sigma^2>0$,
and exact solutions to both block subproblems.
Let $\underline{\bY}^{(t)}$ be optimal for
$\widebreve{\boldsymbol{\Psi}}^{(t-1)}$ and
$L_t=\cL(\widebreve{\boldsymbol{\Psi}}^{(t)})$.
For the full QT objective $f$, the QT identity and block optimality give
\begin{align}
\label{eq:CE_QT_monotonicity}
L_t
&=\max_{\underline{\bY}}
f(\widebreve{\boldsymbol{\Psi}}^{(t)},\underline{\bY})
\geq f(\widebreve{\boldsymbol{\Psi}}^{(t)},\underline{\bY}^{(t)})
\nonumber\\
&\geq
f(\widebreve{\boldsymbol{\Psi}}^{(t-1)},\underline{\bY}^{(t)})
=L_{t-1}.
\end{align}
Since $L_t\leq\sum_k\operatorname{tr}(\tilde{\bC}_k)$,
$L_t$ converges and the surrogate MSE is nonincreasing.
The safeguard retains a quantized candidate only when its actual MSE improves the stored best value.
Thus, the stored MSE is nonincreasing and bounded below.

\section{RIS Beamforming Optimization} \label{sec:BF}
In this section, we investigate the downlink beamforming design for a  discrete-phase RIS-aided MU-MISO system. 
We aim to maximize the sum SE  by jointly optimizing the transmit precoding matrix at the BS and the phase shifts at the RIS.

Let us consider the downlink data transmission where the BS sends independent data streams to $K$ users with precoding. 
Accordingly, the $k$th user's received signal is given by
\begin{align}
    \nonumber 
    y_k &= \left(\bH^{\sf r}_k \bphi\right)^{\sf H}{\bff_k}s_k + \sum_{i\neq k}\left(\bH^{\sf r}_k \bphi\right)^{\sf H}{\bff_i}s_i + n_k,
\end{align}
where $\bff_k \in \bbC^{N}$ is a precoding vector of user $k$ for a data symbol $s_k \sim \cC\cN(0, P)$ and $n_k \sim \cC\cN(0, \sigma^2)$ is additive white Gaussian noise.
The precoder is given by $\bF = [\bff_1, \bff_2, \ldots, \bff_K]\in \bbC^{N\times K}$.
The SE of user $k$ is defined as
\begin{align}
    \label{eq:SE_k}
    R_k(\bF, {\bphi}) = \log_2 \left(1 + \frac{\left|\left(\bH^{\sf r}_k\bphi\right)^{\sf H}{\bff}_k\right|^2}{\sum_{i\neq k}\left|\left(\bH^{\sf r}_k\bphi\right)^{\sf H}{\bff}_i\right|^2 \!+\! \frac{\sigma^2}{P}}\right).
\end{align}
Using \eqref{eq:SE_k}, we formulate the sum SE maximization problem:
\begin{align} 
    \label{eq:main_problem_BF}
    \mathop{{\text{maximize}}}_{\bF, \bphi}& \;\;  \sum_{k = 1}^{K}R_k (\bF, {\bphi})
    \\
    \mathop{{\text{subject to}}}& \;\; {\rm tr}\left(\bF \bF^{\sf H}\right) \leq 1,
    \\
    \label{eq:dis_unitmodulus_constraint}
    & \;\; {\bphi} \in \bbF^{M}.
\end{align}
The joint optimization of $\bF$ and $\bphi$ in \eqref{eq:main_problem_BF}--\eqref{eq:dis_unitmodulus_constraint} is challenging due to the discrete phase-shift constraints as well as the coupled variables, which result in an intractable MINLP. 
To tackle this, we first decouple the variables via an alternating optimization strategy. 
We then leverage our BuLL method in Section~\ref{sec:bussgang} to efficiently resolve the discrete RIS optimization.

\subsection{Discrete RIS Phase-Shift Optimization} \label{subsec:RIS}
For fixed $\bF$, we express the sum SE maximization problem in terms
of the discrete RIS phase shifts as
\begin{align}
    \label{eq:problem_RIS}
    \mathop{{\text{maximize}}}_{{\bphi}}
    &\;\; \sum_{k=1}^{K}R_k(\bphi)
    \\
    \label{eq:const_RIS}
    \mathop{{\text{subject to}}}
    &\;\; \bphi\in\bbF^{M}.
\end{align}
To handle the discrete constraint while accounting for discretization, we apply our BuLL method and introduce a deterministic latent variable $\breve{\bphi}$ through the reference-based working model: $\pmb{\phi}=\beta\breve{\bphi}+\bq$.
We incorporate its average effect by considering the reference-averaged sum SE, where $\bbE_{\bq}[\cdot]$ averages over the discretization distortion:

\begin{align}
    \label{eq:problem_RIS2}
    \mathop{{\text{maximize}}}_{\breve{\bphi}}
    &\;\; \sum_{k=1}^{K}
    \bbE_{\bq}\left[R_k(\beta\breve{\bphi}+\bq)\right].
\end{align}
To obtain a tractable objective using only the reference moments,
we approximate the expected rate by replacing the random signal and
interference powers with their expectations:
\begin{align}
    \nonumber
    \bbE_{\bq}\!\left[R_k(\beta\breve{\bphi}\!+\!\bq)\right]
    &\!\!\stackrel{(a)}{\approx}\!
    \log_2\!\left(\!1 \!+\!
    \frac{
    \bbE_{\bq}\left[
    \left|\left(\bH^{\sf r}_k
    (\beta\breve{\bphi}+\bq)\right)^{\sf H}
    {\bff}_k\right|^2\right]
    }{
    \bbE_{\bq}\!\left[\!
    \sum_{i\neq k}^{K}\!
    \left|\left(\bH^{\sf r}_k
    (\beta\breve{\bphi}\!+\!\bq)\right)^{\sf H}\!
    {\bff}_i\right|^2\!\right]
    \!\!+\!\frac{\sigma^2}{P}
    }\!\right)
    \\
    \label{eq:R_k_bar}
    &=\bar R_k,
\end{align}
where $(a)$ follows the approximation in
\cite{zhang2014power}.
Using $\bbE_{\bq}[\bq]=\mathbf0$ and
$\bbE_{\bq}[\bq\bq^{\sf H}]=\bC_{\bq\bq}$,
the expected powers can be evaluated in closed form, yielding
\begin{align}
    \label{eq:R_k_Cqq}
    \bar{R}_k
    \!=\!\log_2\!\left(\!1\!+\!
    \frac{
    \left|\left(\beta\bH^{\sf r}_k
    \breve{\bphi}\right)^{\sf H}{\bff}_k\right|^2
    +{\bff}_k^{\sf H}\bH^{\sf r}_k\bC_{\bq\bq}
    \left(\bH^{\sf r}_k\right)^{\sf H}{\bff}_k
    }{
    \sum_{i\neq k}\!\left\{
    \left|\left(\beta\bH^{\sf r}_k
    \breve{\bphi}\right)^{\sf H}{\bff}_i\right|^2
    \!\!+\!{\bff}_i^{\sf H}\bH^{\sf r}_k\bC_{\bq\bq}\!
    \left(\bH^{\sf r}_k\right)^{\sf H}{\bff}_i
    \right\}
    \!+\!\frac{\sigma^2}{P}
    }\!\right)\!.
\end{align}
Equation~\eqref{eq:R_k_Cqq} provides a tractable and distortion-aware continuous surrogate using the fixed reference-based moments. 

Using the BuLL framework and averaging over the quantization error, we can utilize well-established optimization methods. 
In particular, we use the GPI approach \cite{choi2019joint} as its power-iteration step enforces a desired structure of phase shifters.
We first transform \eqref{eq:R_k_Cqq} to a  GPI-friendly form.
We fix the latent energy to $\|\breve{\bphi}\|^2=M$
to match the total reference energy and define
$\bw=\breve{\bphi}/\sqrt{M}$, so that $\|\bw\|^2=1$.
Under this normalization, \eqref{eq:R_k_Cqq} is equivalently
expressed as:
\begin{align}
    \label{eq:Rayleigh_quotient}
    \bar{R}_k 
    &= \log_2 \left(\frac{\bw^{\sf H}\bJ_k\bw}{\bw^{\sf H}\bK_k\bw}\right),
\end{align}
where
\begin{align}
    \bJ_k &\!=\! \beta^2 M\!\!\sum_{i=1}^{K}\!\bH_k^{\sf r \sf H}{\bff}_i{\bff}_i^{\sf H}\bH^{\sf r}_k \!+\! \left(\!\sum_{i=1}^{K}{\bff}_i^{\sf H} \bH^{\sf r}_k \bC_{\bq\bq}\bH^{\sf r  \sf H}_k {\bff}_i \!+\! \frac{\sigma^2}{P} \!\right)\!\bI_M,
    \\
    \bK_k &\!=\! \beta^2 M\!\!\sum_{i\neq k}^{K}\!\bH_k^{\sf r \sf H}{\bff}_i{\bff}_i^{\sf H}\bH^{\sf r}_k \!+\! \left(\!\sum_{i\neq k}^{K}{\bff}_i^{\sf H} \bH^{\sf r}_k \bC_{\bq\bq}\bH^{\sf r \sf H}_k {\bff}_i \!+\! \frac{\sigma^2}{P} \!\right)\!\bI_M.
\end{align}
Then,  the sum of the approximations in \eqref{eq:R_k_Cqq} is
\begin{align}
    \label{eq:lamb_obj}
    \cL(\bw) = \log_2 \prod_{k=1}^K \left(\frac{\bw^{\sf H}\bJ_k\bw}{\bw^{\sf H}\bK_k\bw}\right)
    = \log_2 \lambda(\bw).
\end{align}
Finally, we have the reformulated problem {over $\|\bw\|^2=1$} as
\begin{align} 
    \label{eq:AO_RIS_2}
    \mathop{{\text{maximize}}}_{\bw}&\; {\cL}({\bw}).
\end{align}
To solve this problem, we derive Lemma~\ref{lem:NEP_RIS} to find the stationary points of \eqref{eq:AO_RIS_2}.
\begin{lemma}
    \label{lem:NEP_RIS}
The stationary condition of \eqref{eq:AO_RIS_2} is satisfied if the following holds:
\begin{align}
    \label{eq:general_eigen_2}
    \bar{\bK}^{-1}(\bw) \bar{\bJ}(\bw){\bw} = \lambda(\bw)\bw,
\end{align}
where
\begin{align}
    \label{eq:CD_KKT}
    \bar{\bJ}(\bw) \!=\! \lambda_{{\sf 1}}(\bw)\!
    \sum_{k = 1}^{K} \left(\frac{\bJ_k}{{\bw}^{\sf H}\bJ_k {\bw}}\right)\!,\, \bar{\bK}(\bw) \!=\! \lambda_{{\sf 2}}(\bw)\!
    \sum_{k = 1}^{K} \left(\frac{\bK_k}{{\bw}^{\sf H}\bK_k {\bw}}\right).
\end{align}
Here, $\lambda_{\sf{1}}(\bw)$ and $\lambda_{\sf{2}}(\bw)$ can be any  positive functions such that $\lambda(\bw) = \lambda_{\sf{1}}(\bw) / \lambda_{\sf{2}}(\bw)$.
\end{lemma}
\begin{proof}
Please refer to Appendix~\ref{appen:GPI}.
\end{proof}
\begin{algorithm}[t]
    \caption{BuLL GPI-based RIS optimization}
    \label{alg:algorithm_RIS}
    Initialize nonzero $\breve{\bphi}^{(0)}$ and set $\bw^{(0)}=\breve{\bphi}^{(0)}/\|\breve{\bphi}^{(0)}\|$.\;
    \For{$t=1,\ldots,t_{2,\max}$}{
        Build $\bar{\bJ}(\bw^{(t-1)})$ and $\bar{\bK}(\bw^{(t-1)})$ with given $\bF$.\;
        Compute $\bw^{(t)}=\bar{\bK}^{-1}(\bw^{(t-1)})\bar{\bJ}(\bw^{(t-1)})\bw^{(t-1)}$.\;
        Normalize $\bw^{(t)}\leftarrow\bw^{(t)}/\|\bw^{(t)}\|$.\;
        \If{$\|\bw^{(t)}-\bw^{(t-1)}\|\leq\varepsilon_2$}{\textbf{break}.\;}
    }
    \Return{$\breve{\bphi}^{\star}=\sqrt{M}\,\bw^{(t)}$}.
\end{algorithm}
{This stationary condition is a nonlinear generalized eigenvalue condition, since both matrices depend on $\bw$, and motivates the GPI update} \cite{choi2019joint}.
We summarize the resulting GPI-based RIS phase-shift optimization in Algorithm~\ref{alg:algorithm_RIS}.


\begin{remark}
    \normalfont Unlike the previous GPI-based method for RIS optimization \cite{oh2026scalable}, which  introduces a penalty term, i.e.,
    \begin{align}
        \label{eq:mu_search}
        \mu \left(\max_{i=1,\ldots, M}\left\{|\phi_i|^2\right\} - \min_{i=1,\ldots, M}\left\{|\phi_i|^2\right\} \right),
    \end{align}
    into the objective function to handle the unit-modulus constraint, our proposed method eliminates the need for such a penalty term.
    In \cite{oh2026scalable}, determining $\mu$ necessitates an additional line search process with $N_{\mu}$ samples, which incurs extra computational overhead. 
    In contrast, our BuLL method incorporates reference distortion moments into the signal model, allowing Algorithm~\ref{alg:algorithm_RIS} to directly optimize the phase shifts without requiring any penalty terms.
\end{remark}

\subsection{Precoding Design and Alternating Algorithm} \label{subsec:precoder}
In this subsection, we design the precoder $\bF$ while fixing the RIS phase shifts ${\bphi}$.
With fixed ${\bphi}$, the precoder optimization problem is given by
\begin{align} 
    \label{eq:AO_precoder}
    \mathop{{\text{maximize}}}_{\bF}& \; \sum_{k = 1}^{K}\;{R}_{k} \big(\bF \big)
    \\
    \label{eq:power_const2} 
    {\text{subject to}} & \;{\rm tr}\left(\bF \bF^{\sf H}\right) \le 1.
\end{align}
Here, our framework is not restricted to a specific precoder; any suitable scheme can be readily adopted to solve \eqref{eq:AO_precoder}.
Depending on system requirements, any suitable precoding scheme, ranging from iterative algorithms such as weighted minimum mean square error (WMMSE) \cite{shi2011iteratively}  and GPI \cite{choi2019joint} to linear precoders (e.g., zero-forcing), can be seamlessly integrated into this step.
We let $\CMcal{P}(\cdot)$ denote a feasible precoding operator, warm-started from the current precoder.

\begin{algorithm}[t]
    \caption{Alternating beamforming optimization}
    \label{alg:algorithm_AO}
    Define $R_{\Sigma}(\bF,\bphi)=\sum_{k=1}^{K}R_k(\bF,\bphi)$.\;
    Initialize a feasible discrete phase vector $\bphi^{(0)}$ and
    set $\bF^{(0)}\leftarrow\CMcal{P}(\bphi^{(0)})$.\;
    Set $r_0=R_{\Sigma}(\bF^{(0)},\bphi^{(0)})$,
    $r_\star=r_0$, and
    $(\bF^\star,\bphi^\star)=(\bF^{(0)},\bphi^{(0)})$.\;
    \For{$t=1,\ldots,t_{3,\max}$}{
        $\breve{\bphi}_{\mathrm c}\leftarrow$
        Algorithm~\ref{alg:algorithm_RIS}
        $(\bphi^{(t-1)};\bF^{(t-1)})$.\;
        $\bphi^{(t)}\leftarrow\cQ(\breve{\bphi}_{\mathrm c})$ and
        $\bF^{(t)}\leftarrow\CMcal{P}(\bphi^{(t)})$.\;
        Set $r_t=R_{\Sigma}(\bF^{(t)},\bphi^{(t)})$.\;
        \If{$r_t>r_\star$}{
            $(\bF^\star,\bphi^\star)\leftarrow
            (\bF^{(t)},\bphi^{(t)})$ and $r_\star\leftarrow r_t$.\;
        }
        Set $\delta_t=(r_t-r_{t-1})/r_t$.\;
        \If{$0<\delta_t<\varepsilon_3$}{\textbf{break}.\;}
    }
    \Return{$\bF^\star$ {\rm and} $\bphi^\star$}.
\end{algorithm}
Algorithm~\ref{alg:algorithm_AO} uses BuLL-GPI to generate a continuous RIS candidate, which is quantized at every outer iteration.
Here, $\CMcal{P}(\bphi)$ denotes any precoder design method that satisfies the transmit-power constraint.
The resulting phase--precoder pair is evaluated using the actual SE in \eqref{eq:main_problem_BF}.
Although successive updates may decrease the SE, the algorithm separately stores the pair achieving the highest sum SE among all evaluated pairs, including the initial pair, and returns it upon termination.
Thus, the BuLL surrogate guides RIS candidate generation, while the actual SE evaluated with discrete RIS determines which pair is stored.

\subsection{Complexity and Convergence Analysis}
\label{subsec:BF_complexity}
The dense RIS update requires  $\CMcal{O}(T_1M^3)$ complexity, where $T_1$ denotes the iteration count for Algorithm~\ref{alg:algorithm_RIS}.
The precoding update costs $\CMcal{O}(T_2I_P)$, where $T_2$ and $I_P$ denote its iteration count and per-iteration cost.
Including matrix construction and initialization, a dense implementation has complexity $\CMcal{O}(T_{3}(T_1 M^3 + T_2 I_P))$ where $T_3$ is the number of outer iterations.
Our implementation exploits the low-rank structure of the GPI linear system to avoid an explicit $M\times M$ matrix inversion.
The method avoids enumerating all $2^{bM}$ discrete phase vectors and requires no penalty-parameter tuning.

Convergence results for GPI~\cite{yoo2026scalable} and precoding methods such as WMMSE~\cite{shi2011iteratively} are established under their respective assumptions for their individual objectives.
These results do not directly establish convergence of Algorithm~\ref{alg:algorithm_AO}, since the RIS update in Algorithm~\ref{alg:algorithm_RIS} optimizes a distortion-aware surrogate, whereas the precoder update targets the actual SE.
Consequently, alternating these updates need not improve the actual SE  particularly after phase quantization.
To ensure monotonicity of the retained performance, the safeguard evaluates each quantized candidate pair using the actual SE and replaces the stored pair only when an improvement is obtained.

Let $(\bF^{(s)},\bphi^{(s)})$ denote the evaluated feasible
candidate pairs, with $s=0$ corresponding to initialization,
and define
\[
r_\star^{(t)}
=\max_{0\leq s\leq t}
R_{\Sigma}(\bF^{(s)},\bphi^{(s)}).
\]
By construction, $r_\star^{(t)}\geq r_\star^{(t-1)}$.
For fixed channels, bounded transmit power, and positive noise variance, these values are bounded above and hence converge if the iterations are continued indefinitely.
The returned pair performs no worse than the initial pair.

\section{Numerical Results} \label{sec:numerical}



\subsection{Simulation Environments}
In the considered system, we assume that both antennas at the BS and elements of the RIS are arranged in uniform planar arrays (UPAs).
High-frequency channels typically exhibit a sparse multipath structure due to limited diffraction and are modeled using the geometric channel model:
\begin{align}
    \label{eq:H_1}
    \bH_{1} = \frac{1}{\sqrt{L_{\sf BR}}} \sum_{i=1}^{L_{\sf BR}} \sqrt{\gamma_{1}} \alpha_{1,i} \ba_{\sf B}(\vartheta^{a}_{i}, \vartheta^{e}_{i}) \ba_{\sf R}^{\sf H} (\varphi^{a}_{i}, \varphi^{e}_{i}),
\end{align}
where $\gamma_{1}$ is the path gain of the BS-RIS link, $\alpha_{1,i} \sim \cC \cN (0,1)$ is small-scale fading, $\vartheta^{a}_{i}$ and $\vartheta^{e}_{i}$ denote the azimuth and elevation angles of departure (AoD) at the BS, $\varphi^{a}_{i}$ and $\varphi^{e}_{i}$ denote the corresponding angles of arrival (AoA) at the RIS, and $\ba(\cdot)$ denotes an array response vector (ARV).

Let $N_y$ and $N_z$ denote the number of horizontal and vertical BS antenna elements, i.e., $N = N_y \times N_z$. 
The ARVs at the BS and RIS are defined as
\begin{align}
    \ba_{\sf B}(\vartheta^{a}_{i}, \vartheta^{e}_{i}) &= \ba_{\sf B}^{y}(\vartheta^{a}_{i}, \vartheta^{e}_{i}) \otimes \ba_{\sf B}^{z}(\vartheta^{e}_{i}),
    \\
    \ba_{\sf R}(\varphi^{a}_{i}, \varphi^{e}_{i}) &= \ba_{\sf R}^{y}(\varphi^{a}_{i}, \varphi^{e}_{i}) \otimes \ba_{\sf R}^{z}(\varphi^{e}_{i}),
\end{align}
where
\begin{align}
    \label{eq:a_B_y}
    &\ba_{\sf B}^{y}(\vartheta^{a}_{i}, \vartheta^{e}_{i}) \!
    \\ \nonumber
    &=\! \left[\!1, e^{j 2\pi \frac{\Delta^y_{\sf B}}{\lambda_c}\!\cos(\vartheta^{e}_{i})\!\sin(\vartheta^{a}_{i})},\ldots,e^{j 2\pi (N_y - 1)  \frac{\Delta^y_{\sf B}}{\lambda_c}\!\cos(\vartheta^{e}_{i})\!\sin(\vartheta^{a}_{i})}\!\right]^{\!\sf T}\!\!\!,
    \\
    \label{eq:a_B_z}
    &\ba_{\sf B}^{z}(\vartheta^{e}_{i}) \!=\! \left[1, e^{j 2\pi \frac{\Delta^z_{\sf B}}{\lambda_c}\!\sin(\vartheta^{e}_{i})}, \!\ldots\!, e^{j 2\pi (N_z - 1)  \frac{\Delta^z_{\sf B}}{\lambda_c}\!\sin(\vartheta^{e}_{i})} \right]^{\sf T}\!\!\!,
\end{align}
and $\Delta^y_{\sf B}$, $\Delta^z_{\sf B}$ are the distances between adjacent BS antennas along the two axes.
Similarly, for the RIS with $M_y$ horizontal and $M_z$ vertical elements ($M = M_y \times M_z$), the ARVs $\ba_{\sf R}^{y}(\varphi^{a}_{i}, \varphi^{e}_{i})$ and $\ba_{\sf R}^{z}(\varphi^{e}_{i})$ are defined analogously to \eqref{eq:a_B_y} and \eqref{eq:a_B_z}, respectively.
We assume that the total number of spatial paths is $L_{\sf RU}$ for all RIS-user links. 
The RIS-user channel is 
\begin{align}
    \label{eq:h_2}
    \bh_{2,k} = \frac{1}{\sqrt{L_{\sf RU}}} \sum_{i=1}^{L_{\sf RU}} \sqrt{{\gamma}_{2,k}} \alpha_{2, k,i} \ba_{\sf R}(\vartheta^{{\sf R},a}_{k,i}, \vartheta^{{\sf R},e}_{k,i}),
\end{align}
where $\gamma_{2, k}$ is the path gain of the RIS-user link, $\alpha_{2,k,i} \sim \cC \cN (0,1)$ is small-scale fading, and $(\vartheta^{{\sf R},a}_{k,i}, \vartheta^{{\sf R},e}_{k,i})$ is the azimuth and elevation AoDs from the RIS to user $k$ for the $i$th path.

Based on \cite{akdeniz2014millimeter}, we adopt the following pathloss model:
\begin{align}
    \label{eq:PL}
    {\rm PL}(d)= \alpha_{\sf PL} + \beta_{\sf PL} 10 \log_{10} d + \chi \ [{\rm dB}],
\end{align}
where $d$ denotes the link distance in meters and $\chi \sim \cN (0,\sigma_s^2)$ is log-normal shadowing.
This model is applied to the path gain terms in \eqref{eq:H_1} and \eqref{eq:h_2} using the linear gain $\gamma(d)=10^{-{\rm PL}(d)/10}$.
Following the experimental data in \cite{akdeniz2014millimeter} for 28 GHz channels, we set $\alpha_{\sf PL} = 61.4\;{\rm dB}$, $\beta_{\sf PL}=2$, and $\sigma_s = 5.8\;{\rm dB}$.
The noise power is given by
\begin{align}
    \label{eq:P_noise}
      P_{\sf noise,dBm}  = -174 + 10 \log_{10} W + n_f \ [{\rm dBm}],
\end{align}
where $W$ and $n_f$ are the channel bandwidth and noise figure at the BS.
Here, $W$ is expressed in hertz, and the linear noise power is $\sigma^2=10^{(P_{\sf noise,dBm}-30)/10}$ in watts.
We set $W = 500$ MHz and  the noise figure to $n_f=5$ dB.
In the considered channel model, we set $\Delta_{\sf B}^{y}/\lambda_c \!=\! \Delta^{z}_{\sf B}/\lambda_c \!=\! \Delta_{\sf R}^{y}/\lambda_c \!=\! \Delta_{\sf R}^{z}/\lambda_c \!=\! 0.5$, and randomly generate the azimuth and elevation angles as $(\vartheta^{a}_{i}, \varphi^{a}_{i}, \vartheta^{{\sf R}, a}_{k,i}) \in [-\pi, \pi)$ and $(\vartheta^{e}_{i}, \varphi^{e}_{i}, \vartheta^{{\sf R}, e}_{k,i}) \in [-\pi/2, \pi/2],\forall i, k$.

\begin{figure}[t]    
    {\centerline{\resizebox{0.71\columnwidth}{!}{\includegraphics{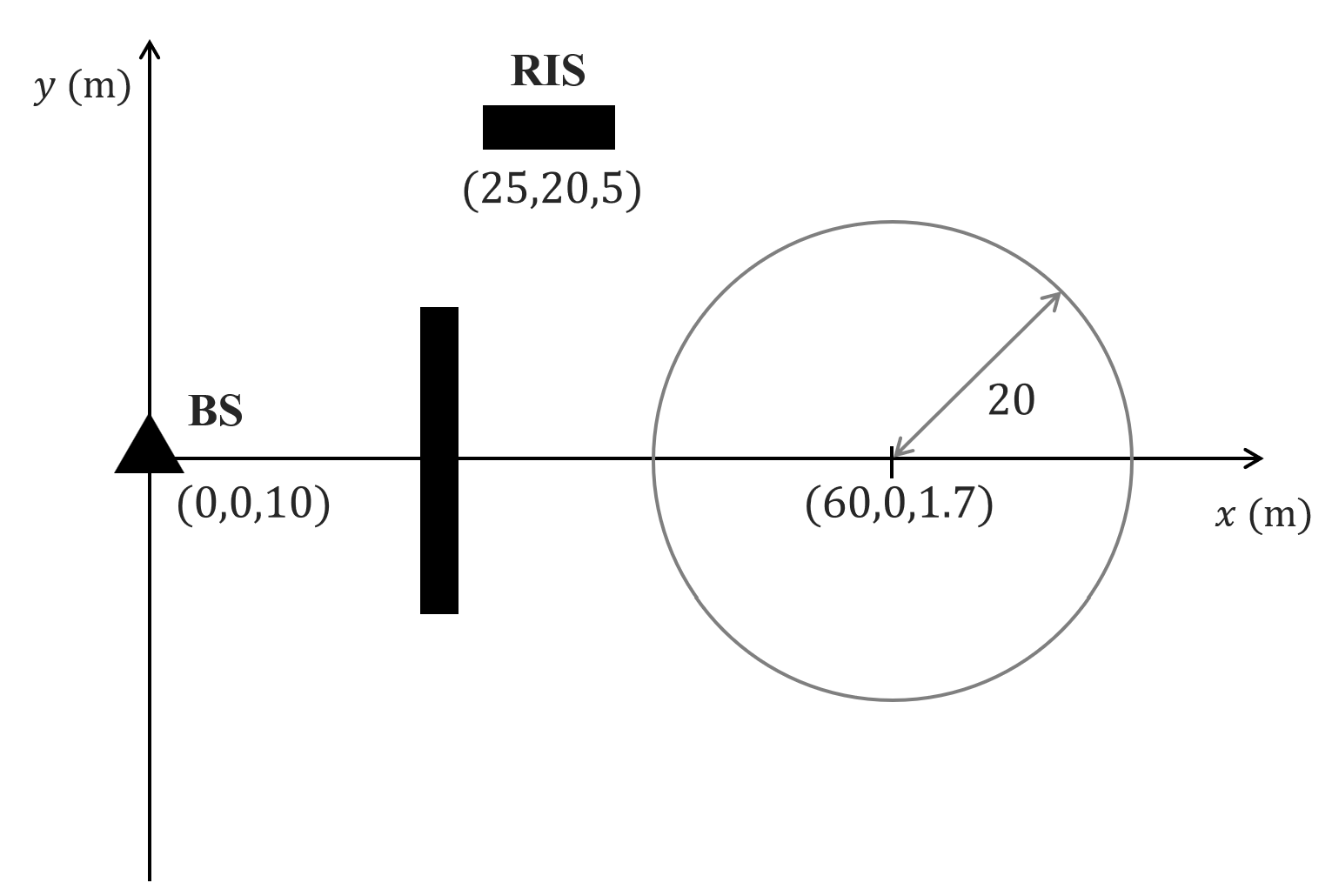}}}}
    \vspace{-0.8em}
    \caption{A schematic of the multiuser RIS-aided network}
    \label{fig:network}
\end{figure}
In our simulations, a three-dimensional Cartesian coordinate system is adopted to define the spatial locations of the network entities.
Fig.~\ref{fig:network} presents a top-down view of the simulation setup projected onto the $xy$-plane.
Specifically, the BS and the RIS are located at $(0, 0, 10)$ m and $(25, 20, 5)$ m, respectively. 
The users are randomly distributed within a circular region of radius 20 m at a height of 1.7 m. 
The center of this circular region is set to $(60, 0, 1.7)$.
Furthermore, the number of spatial paths is set to $L_{\sf BR} = L_{\sf RU} = 8$.

\subsection{Validation of the BuLL Metrics and Moments.}
\label{subsec:surrogate_validation}
\begin{figure}[t]    
    {\centerline{\resizebox{0.85\columnwidth}{!}{\includegraphics{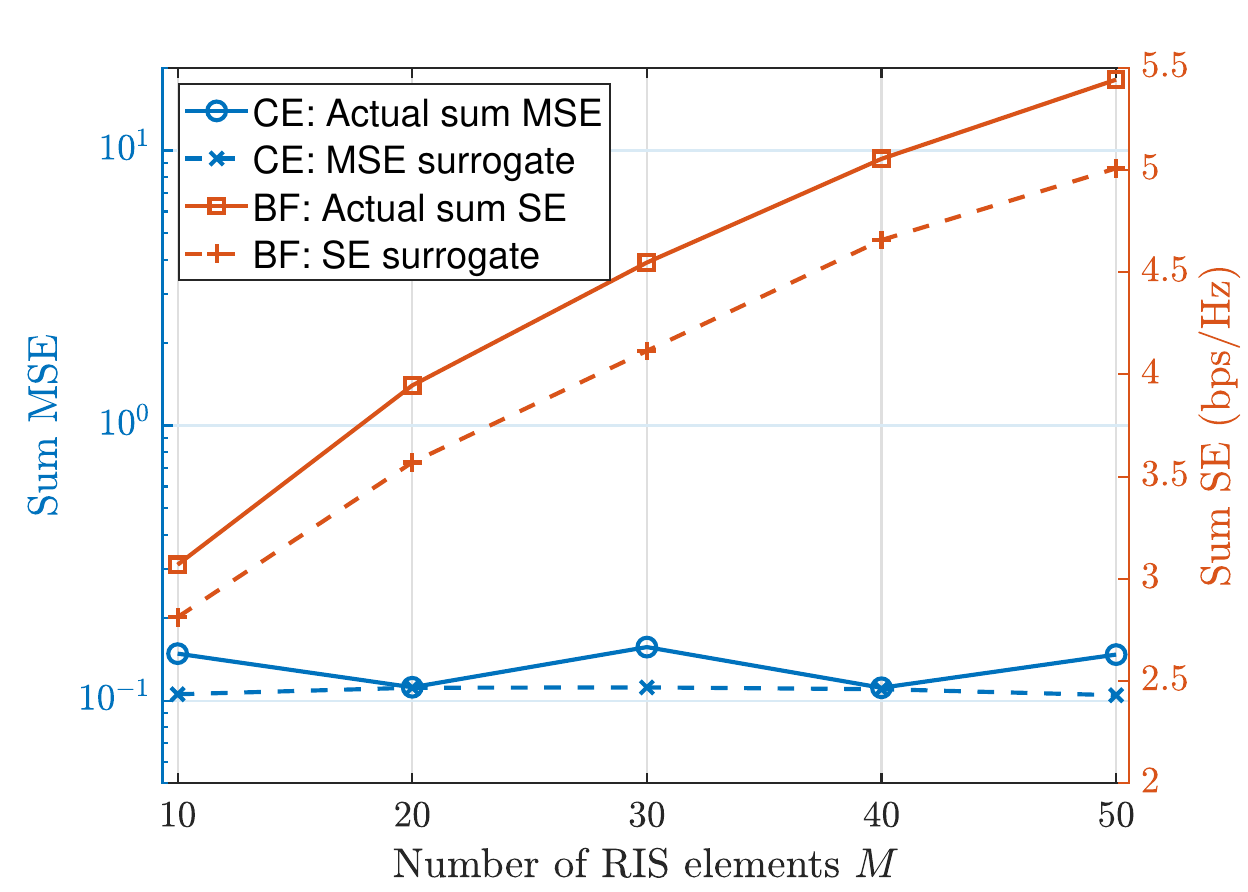}}}
    \vspace{-0.8em}
    \caption{The actual and surrogate metrics versus the number of RIS elements $M$: sum MSE (left axis) and sum SE (right axis).}
    \label{fig:surrogate_validation}} 
\end{figure}

To assess how well the proposed surrogates represent the actual
quantized-design metrics, we compare the sum MSE and sum SE as the
number of RIS elements $M$ varies for $b=1$ bit.
For channel estimation, the Jensen-derived surrogate in \eqref{eq:LB_Cov} is compared with the actual MSE evaluated using the corresponding quantized pattern.
For beamforming, the RIS latent vector is optimized with a fixed normalized
random precoder, which is used in both the surrogate in \eqref{eq:R_k_Cqq} and actual quantized-phase SE evaluations.

As shown in Fig.~\ref{fig:surrogate_validation}, the Jensen surrogate
provides MSE values comparable to the actual metric, although the
latter exhibits more variation with $M$.
The SE surrogate closely follows the increasing trend of the actual
sum SE, with a remaining approximation gap.
These results illustrate the empirical correspondence between the
surrogates and the implemented metrics under the considered settings.

\begin{table}[t]
\centering
\caption{Comparison of Reference and Empirical Bussgang Moments.}
\vspace{-0.2cm}
\label{tab:fixed_bussgang_moments}
\small
\renewcommand{\arraystretch}{1.12}
\setlength{\tabcolsep}{3pt}
\begin{tabular}{c|c|ccc}
\hline \hline
\textbf{Metric} & $\boldsymbol{b}$ &
\textbf{Reference} & \textbf{Gaussian} & \textbf{Optimized} \\
\hline

\multirow[c]{2}{*}{$\beta$}
& 1 & 0.5642 & 0.5642 & 0.5891 \\
& 2 & 0.7979 & 0.7979 & 0.8064 \\
\hline

\multirow[c]{2}{*}{$\delta_{\rm auto}$}
& 1 & 0 & 0.0497 & 0.1873 \\
& 2 & 0 & 0.0498 & 0.0618 \\
\hline

\multirow[c]{2}{*}{$\delta_{\rm cross}$}
& 1 & 0 & 0.0499 & 0.0722 \\
& 2 & 0 & 0.0500 & 0.0608 \\
\hline \hline
\end{tabular}
\end{table}

Furthermore, we evaluate whether the fixed Gaussian-reference Bussgang moments remain representative after beamforming optimization by collecting the final continuous RIS latent vectors $\breve{\bphi}_r^\star$ before phase quantization.
Their actual quantization residuals are defined as
\begin{align}
\bq_r \triangleq \cQ(\breve{\bphi}_r^\star)
-\beta_{\rm ref}\breve{\bphi}_r^\star,
\end{align}
where $\beta_{\rm ref}$ is fixed before optimization.
The empirical gain is fitted separately as
\begin{align}
    \widehat{\beta} = \frac{\sum_{r=1}^{R}\breve{\bphi}_r^{\star {\sf H}}\cQ(\breve{\bphi}_r^\star)
    }{\sum_{r=1}^{R}\|\breve{\bphi}_r^\star\|^2}.
\end{align}
In addition, we report its real part together with the normalized auto-covariance and cross-covariance metrics
\begin{align}
\delta_{\rm auto}
=
\frac{
\big\|\widehat{\mathbf C}_{\bq\bq}
-\mathbf C_{\bq\bq}^{\rm ref}\big\|_F
}{\big\|\bC_{\bq\bq}^{\rm ref}\big\|_F
},\;\;
\delta_{\rm cross}
=
\frac{
\big\|\widehat{\mathbf C}_{\bq\breve{\bphi}}\big\|_F
}{\sqrt{\big\|\bC_{\bq\bq}^{\rm ref}\big\|_F \big\|\bC_{\breve{\bphi}\breve{\bphi}}^{\rm ref}\big\|_F}},
\end{align}
where $\mathbf C_{\bq\bq}^{\rm ref}\triangleq(1-\beta_{\rm ref}^{2})\mathbf I_M$ and $\mathbf C_{\breve{\bphi}\breve{\bphi}}^{\rm ref}\triangleq\mathbf I_M$, and also $\widehat{\mathbf C}_{\bq\bq}$ and $\widehat{\mathbf C}_{\bq\breve{\bphi}}$ denote the empirical auto- and cross-covariance matrices.

Table~\ref{tab:fixed_bussgang_moments} compares post-optimization moments with analytical and reference benchmarks.
Each Gaussian benchmark is the median of the corresponding empirical metric over independently generated ensembles with $M=50$ and $R=10^4$ as the optimized ensemble.
For both phase resolutions, the empirical gain closely matches its analytical value, and the normalized cross-covariance and auto-covariance error magnitude are comparable to the Gaussian benchmarks.
These results provide empirical support for using fixed Gaussian-reference moments in the proposed distortion-aware surrogates.

\subsection{Channel Estimation Performance Evaluation}
\label{subsec:CE_evalution}
In this subsection, we evaluate the channel estimation performance in terms of normalized MSE (NMSE) defined as 
\begin{align}
    {\rm NMSE}_k = 
    {\rm tr}\big(\bR^{\sf e}_k\big)\;/\;{\rm tr}\left(\bC_k\right),\; \forall k  
\end{align}
for $N=8$, $K = 4$, and $b=1$ bit.
For comparison, we consider the following schemes:
\begin{itemize}
    \item {\textbf{BuLL-QT}:} The proposed algorithm in Algorithm~\ref{alg:algorithm_QT}.

    \item {\textbf{QT}:} Algorithm~\ref{alg:algorithm_QT} with continuous relaxation of the discrete constraint, i.e., without accounting for BuLL effects.

    \item {\textbf{DFT $(b=\infty)$}:}  A continuous-phase DFT pattern is used as an optimal-performance benchmark.

    \item {\textbf{CE-Baseline}:} A discrete RIS beam pattern design presented in \cite{you2020channel} based on a quantized DFT and truncated Hadamard method. 

    \item {\textbf{Random}:} The RIS beam pattern consists of randomly generated  phase shifts.
\end{itemize}

\begin{figure}[t]    
    {\centerline{\resizebox{0.85\columnwidth}{!}{\includegraphics{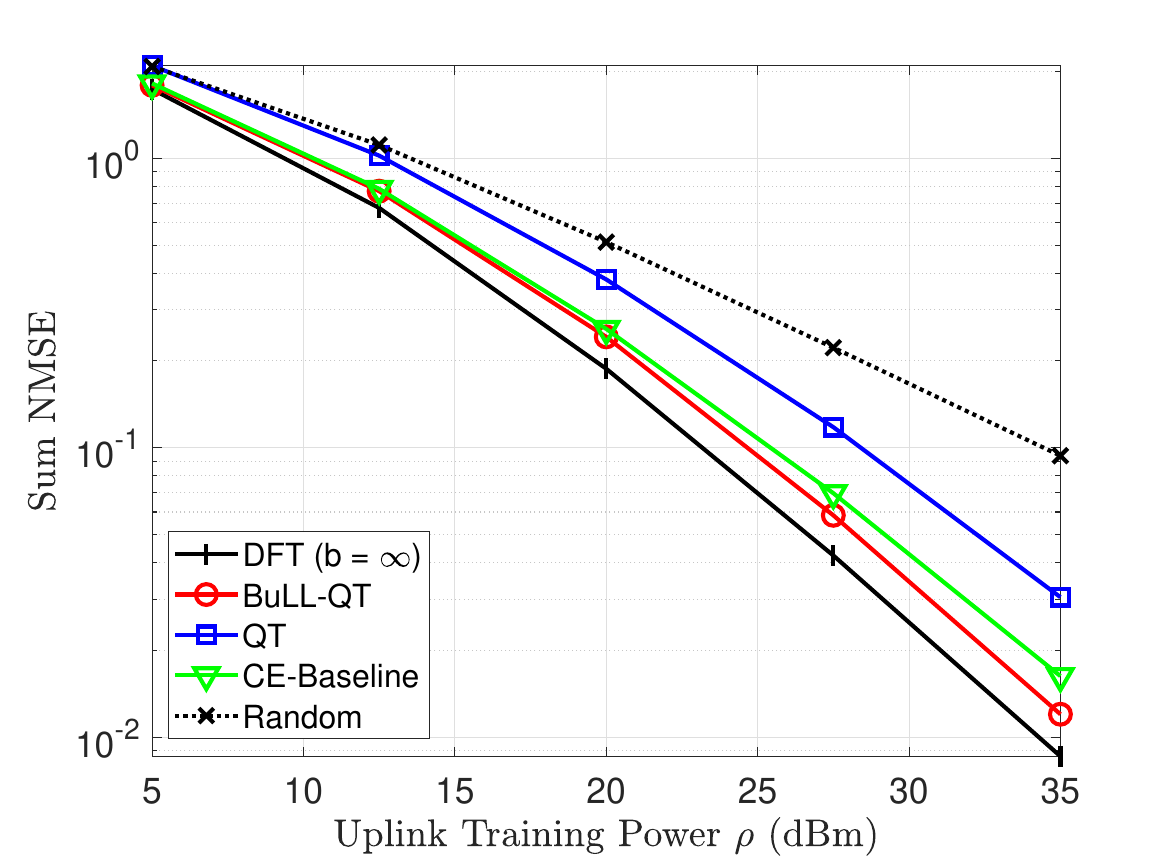}}}
    \vspace{-0.8em}
    \caption{The sum NMSE versus the uplink training power $\rho$ for $N = 8$ BS antennas, $K = 4$ users, and $M = 25$ RIS elements.}
    \label{fig:rhovsMSE}} 
\end{figure}
Fig.~\ref{fig:rhovsMSE} shows the NMSE performance with respect to the uplink training power $\rho$ for $M=25$.
The proposed BuLL-QT closely approaches the continuous-phase benchmark and achieves lower NMSE than QT.
Its relative NMSE advantage over QT becomes more pronounced as the training power increases.
These results support the benefit of incorporating reference-averaged discretization effects into the continuous objective, whereas QT optimizes a continuous relaxation without explicitly accounting for subsequent phase quantization.

\begin{figure}[t]    
    {\centerline{\resizebox{0.85\columnwidth}{!}{\includegraphics{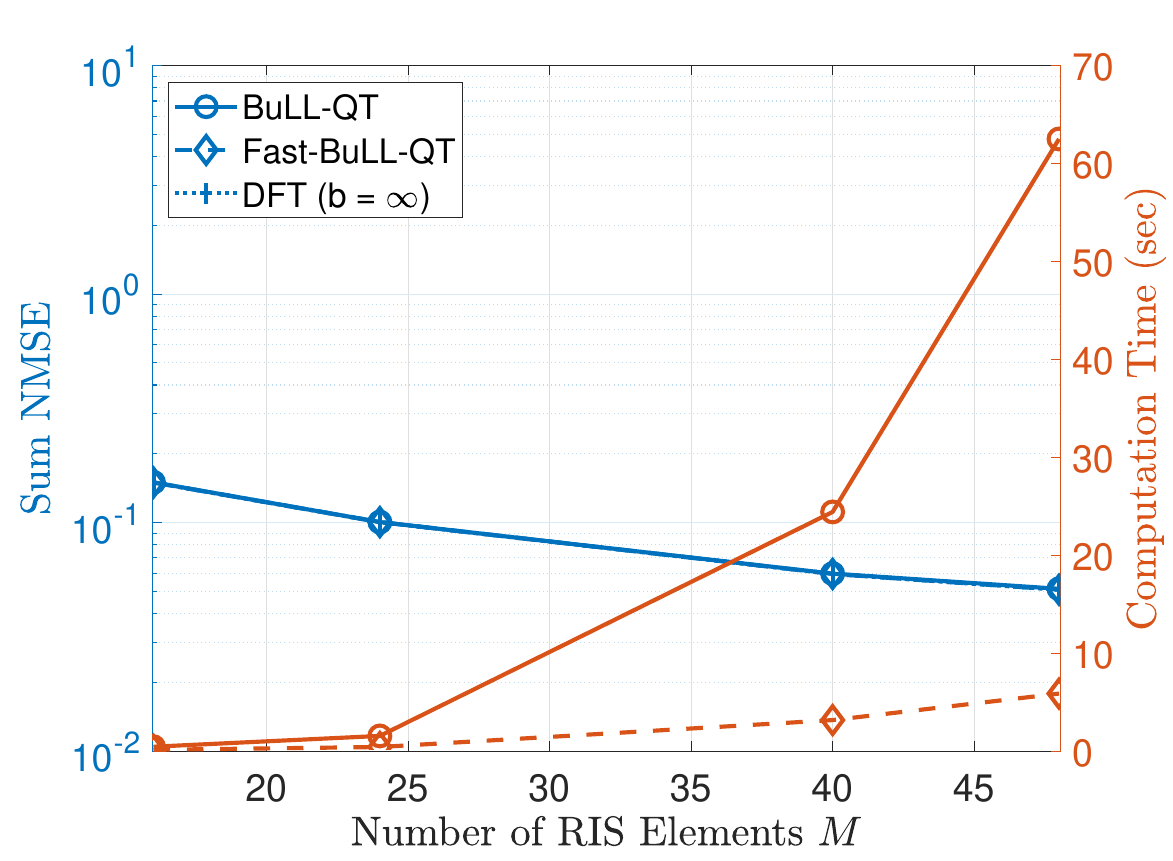}}}
    \vspace{-0.8em}
    \caption{The sum NMSE (left axis) and computation time  (right axis) versus the number of RIS elements $M$ for $N=8$ BS antennas, $K=4$ users, and $\rho=23$ dBm.}
    \label{fig:MvsMSE}} 
\end{figure}
Fig.~\ref{fig:MvsMSE} presents the NMSE performance and computation time versus the number of RIS elements $M$ for $\rho=23~{\rm dBm}$.
At $M\in\{16,24,40,48\}$ with $T=M$, the available Hadamard patterns satisfy $\boldsymbol{\Psi}^{\sf H}\boldsymbol{\Psi}=M\bI_M$ and yield the same LMMSE performance as the continuous-phase DFT benchmark.
The simulation results show that BuLL-QT and its accelerated implementation closely track the benchmark performance across all evaluated values of $M$.
The computation times are measured in MATLAB on a workstation with an Intel i9-13900K CPU and 64~GB RAM.
The computation-time gap between BuLL-QT and Fast-BuLL-QT widens significantly as $M$ increases.
This computational advantage comes from the inversion-free latent updates of Fast-BuLL-QT described in Section~\ref{subsec:CE_complexity}.

\subsection{Beamforming Performance Evaluation}
\label{subsec:BF_evalution}
In this subsection, we evaluate the sum SE performance of the proposed beamforming algorithm  (Algorithm~\ref{alg:algorithm_AO}) for $N=16$, $K = 4$, and $b=1$ bit.
For comparison, we use the WMMSE algorithm  \cite{shi2011iteratively} for precoding at the BS $\CMcal{P}(\cdot)$ for all cases:
\begin{itemize}
    \item {\textbf{BuLL-GPI}:} 
    The proposed alternating algorithm in Algorithm~\ref{alg:algorithm_AO}, using BuLL-GPI for the RIS update.

    \item {\textbf{GPI $(b=\infty)$}:} The GPI-based phase-shift optimization algorithm \cite{oh2026scalable} with $b=\infty$.

    \item {\textbf{GPI}:} The GPI-based phase-shift optimization algorithm \cite{oh2026scalable} with continuous relaxation.
    The parameter for the penalty term is obtained by a line search with $N_{\mu}=5$.

    \item {{\textbf{BF-Baseline}:} A discrete RIS-aided beamforming design based on a successive refinement approach \cite{wu2019beamforming}.}

    \item {\textbf{Random}:} RIS phase shifts are randomly generated.
\end{itemize}

\begin{figure}[t]    
    {\centerline{\resizebox{0.85\columnwidth}{!}{\includegraphics{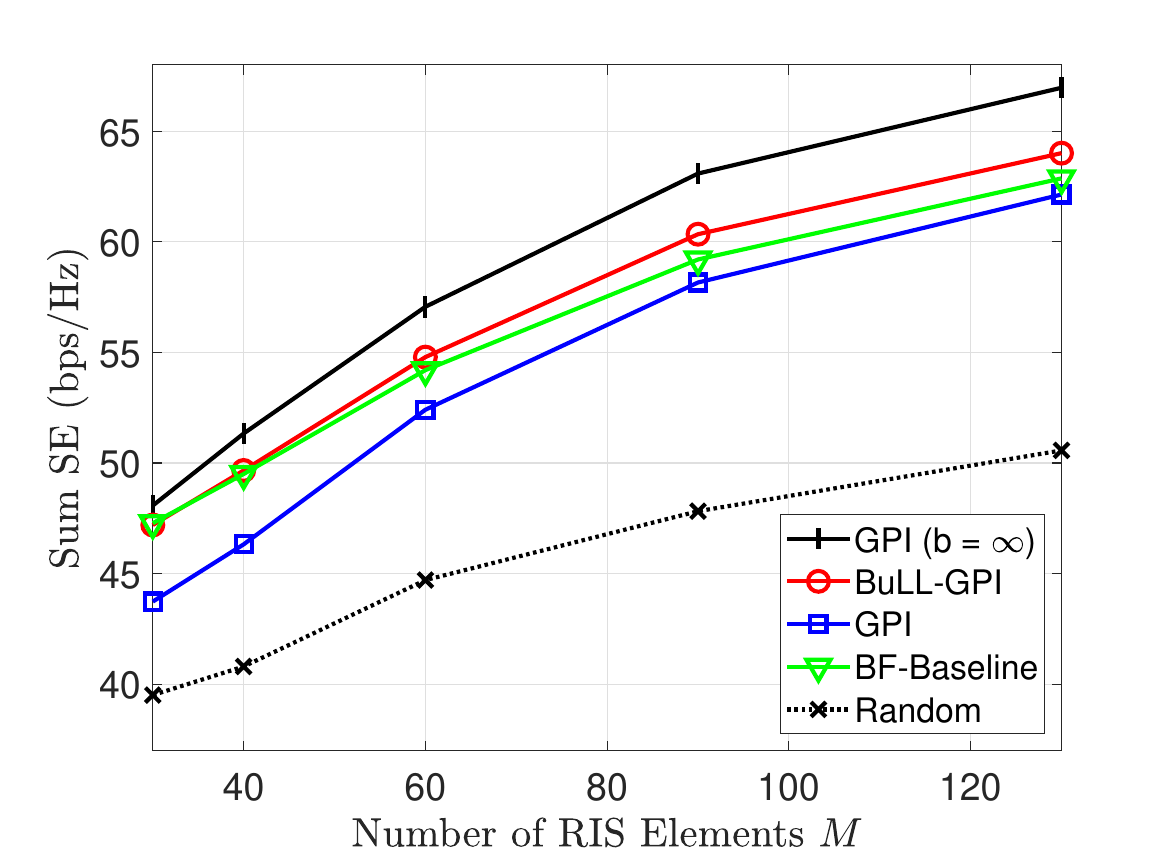}}}
    \vspace{-0.8em}
    \caption{{\color{black}The sum SE versus the number of RIS elements $M$ for $N = 16$ BS antennas, $K = 4$ users, and $P = 40$ dBm.}}
    \label{fig:MvsSE}} 
\end{figure}
\begin{table}[t]
\renewcommand{\arraystretch}{1.3}
\caption{{\color{black}Computation Time and Sum SE  Comparisons}}
\vspace{-0.8em}
\label{tab:comp_time}
\centering
\color{black}
\begin{tabular}{l c c}
\hline\hline
\textbf{Scheme} & \textbf{CPU Time (msec)} & \textbf{Sum SE (bps/Hz)} \\
\hline
BuLL-GPI            & {\bf32.18}   & {\bf 28.83} \\
BF-Baseline   & 238.50  & 28.68 \\
GPI with $N_{\mu} = 1$  & {35.35} & 26.02 \\
GPI with $N_{\mu} = 5$  & 160.99    & 27.26 \\
\hline\hline
\end{tabular}
\end{table}
In Fig.~\ref{fig:MvsSE}, we evaluate the sum SE with respect to the number of RIS elements $M$ for $P=40$ dBm.
The sum SE for all schemes increases with $M$ due to the enhanced passive beamforming gain.
As shown in Fig.~\ref{fig:MvsSE}, BuLL-GPI maintains its advantage over the conventional GPI across all values of $M$, further confirming the effectiveness of our BuLL method.
While BF-Baseline achieves comparable SE at small $M$, its gap with BuLL-GPI widens as $M$ increases, since its element-wise sequential updates become increasingly susceptible to local optima as inter-element coupling intensifies.
These results verify that BuLL-GPI offers superior scalability for large-scale RIS deployments, where jointly accounting for all RIS elements becomes critical.

In Table~\ref{tab:comp_time}, we assess the considered methods for $M=60$ and $P=20$ dBm.
Since BuLL-GPI eliminates the need for penalty parameter tuning, it achieves the lowest computation time among all schemes while attaining the highest sum SE.
In contrast, the conventional GPI-based RIS method requires a line search over the penalty parameter $\mu$ with $N_\mu$ candidates to enforce the unit-modulus constraint.
As shown in the table, increasing $N_\mu$ from 1 to 5 improves SE performance but at the cost of proportionally higher computation time, revealing a fundamental trade-off between performance and complexity.
In addition, although BF-Baseline achieves a comparable SE to BuLL-GPI, it relies on a successive refinement strategy that sequentially updates each discrete variable while fixing the others, incurring significantly higher computational cost.

\begin{figure}[t]    
    {\centerline{\resizebox{0.85\columnwidth}{!}{\includegraphics{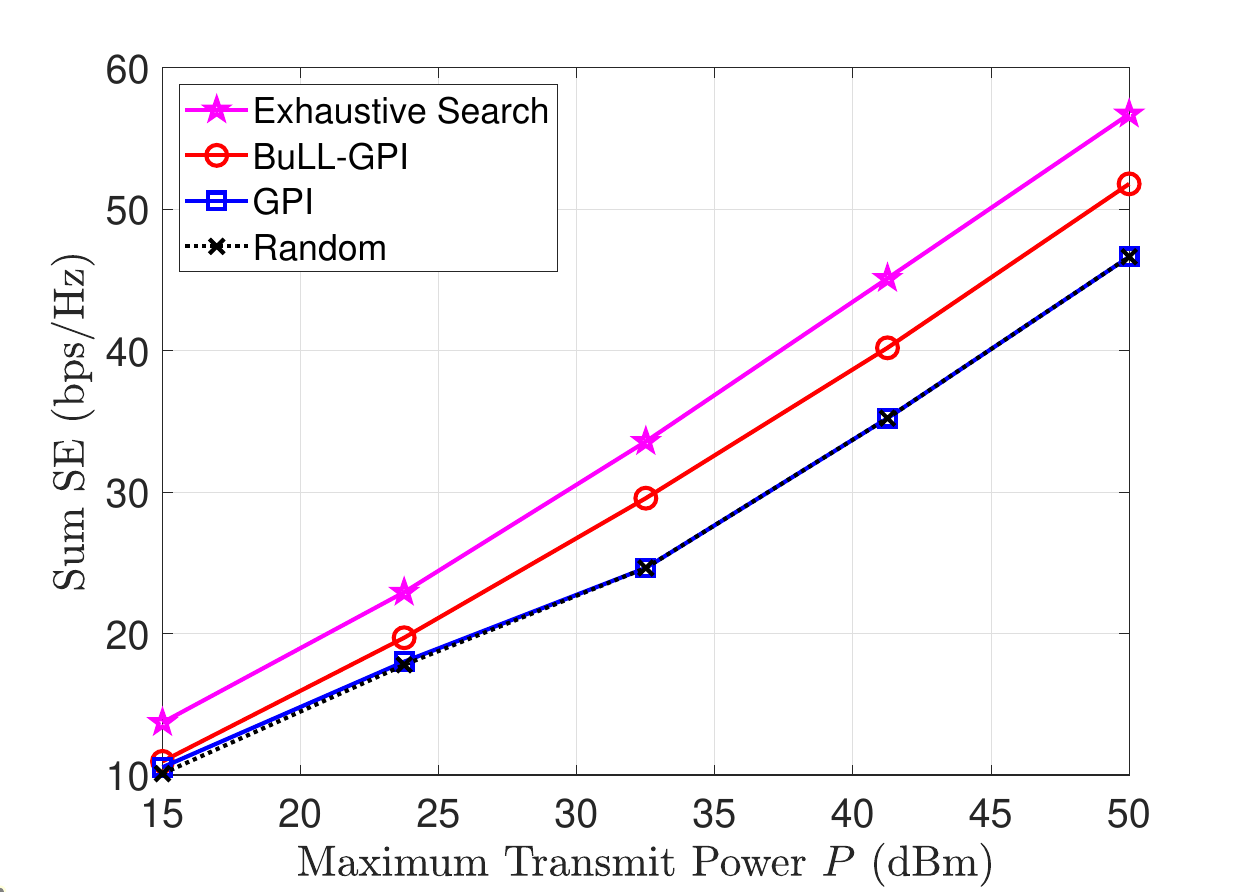}}}
    \vspace{-0.5em}
    \caption{The sum SE versus the maximum transmit power $P$ dBm under small RIS elements $M=12$ for $N = 16$ BS antennas and $K = 4$ users.}
    \label{fig:optimality}} 
\end{figure}

To assess the solution quality of BuLL-GPI,
Fig.~\ref{fig:optimality} presents the sum SE versus the maximum
transmit power $P$ for $N=16$, $K=4$, and $M=12$.
This small-$M$ setting makes exhaustive search
of the discrete RIS configurations computationally feasible,
allowing comparison with an optimal benchmark.
BuLL-GPI consistently outperforms GPI and achieves
sum SE closer to this benchmark across the evaluated power range.
These results support the effectiveness of our BuLL method with the distortion-aware surrogate for generating high-quality discrete RIS designs without exhaustive phase enumeration.

\begin{figure}[t]    
    {\centerline{\resizebox{0.85\columnwidth}{!}{\includegraphics{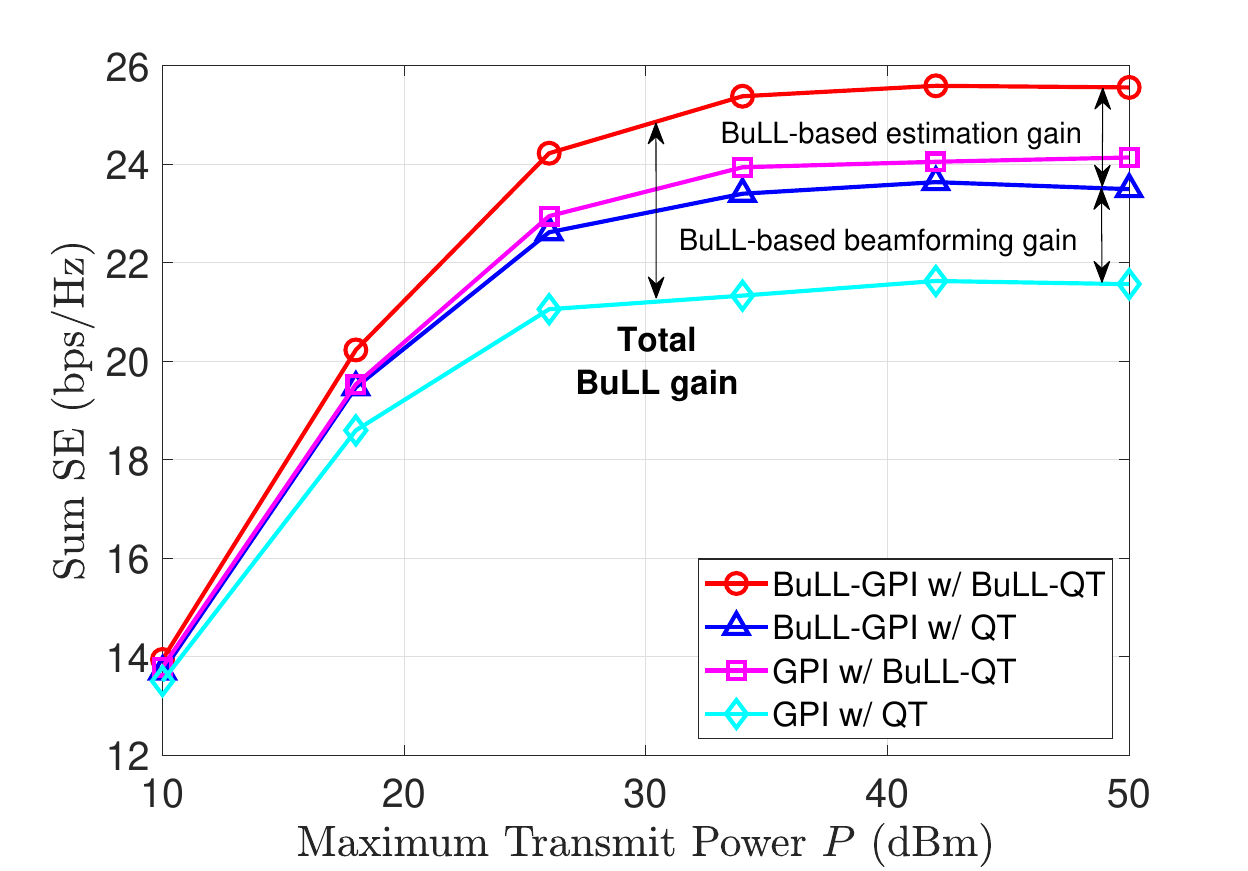}}}
    \vspace{-0.7em}
    \caption{The sum SE versus the maximum transmit power $P$ dBm for $N = 16$ BS antennas, $K = 4$ users, and $M = 30$ RIS elements with imperfect CSIT from channel estimation.}
    \label{fig:PvsSE_imperfect}} 
\end{figure}

\subsection{Performance of Precoding with Channel Estimation}
We now evaluate the beamforming algorithms under imperfect CSIT by incorporating the MSE resulting from the channel estimation algorithms.
To this end, we generate cascaded channel error vectors that preserve the second-order statistics of the LMMSE estimation error, which is the key quantity minimized by the MSE-based RIS pattern design in \eqref{eq:main_problem}.
Specifically, letting $\bR^{\sf e}_k = \bU^{\sf e}_k \boldsymbol{\Lambda}^{\sf e}_k \bU^{\sf e\,{\sf H}}_k$ denote the eigendecomposition of the error covariance matrix, we construct $\be_k = \bU^{\sf e}_k (\boldsymbol{\Lambda}_k^{\sf e})^{\frac{1}{2}}\,\bz_k$ with $\bz_k \sim \cC \cN({\bf 0}, \bI_{NM})$, so that $\bbE[\be_k\be_k^{\sf H}]=\bR^{\sf e}_k$, which yields the user-wise MSE, i.e.,  ${\rm MSE}_k = {\rm tr}(\bR^{\sf e}_k)$ for all $k$.
The error covariance matrices are computed via the LMMSE estimator with the RIS pattern designed by each algorithm.

Fig.~\ref{fig:PvsSE_imperfect} shows  the sum SE versus the maximum transmit power $P$ for $N=16$, $K=4$, and $M=30$.
To isolate the individual contributions of the BuLL method, we consider four combinations of channel estimation (BuLL-QT vs.\ QT) and beamforming (BuLL-GPI vs.\ GPI) schemes.
It is important to emphasize that each pair employs an \emph{identical} optimization algorithm; the \emph{sole} difference is whether the BuLL method is incorporated.
Therefore, every performance gap observed in Fig.~\ref{fig:PvsSE_imperfect} is entirely attributed to our BuLL method in discrete optimization.
With a fixed beamforming method, replacing QT with BuLL-QT yields a substantial estimation gain, while with fixed channel estimation, employing BuLL-GPI provides a distinct beamforming gain over GPI.
These two gains are complementary: the fully BuLL-based pipeline (BuLL-GPI with BuLL-QT) achieves a significant total gain over its counterpart without the proposed approach (GPI with QT).
These results confirm that our BuLL method serves as a plug-in enhancement that can be readily embedded into existing algorithms to unlock substantial performance improvements in various discrete optimization problems.

\section{Conclusion} \label{sec:conclusion}
In this paper, we proposed the BuLL-based discrete optimization approach and applied it to discrete RIS-aided multiuser communication systems.
The core idea is to introduce a continuous latent variable and characterize its statistical relationship with the discrete variable through the Bussgang decomposition under a Gaussian reference distribution. The resulting Bussgang moments are used to construct a distortion-aware continuous surrogate for the discrete optimization problem.
A distinct advantage of BuLL is its ability to account for discretization errors within the surrogate without relying on penalty terms.
Using this approach, we developed BuLL-QT for uplink RIS pattern design and BuLL-GPI for downlink RIS beamforming optimization.
Numerical results demonstrated competitive channel estimation
and SE performance for BuLL-QT and BuLL-GPI, respectively.
These findings confirm that incorporating our BuLL method offers a principled and efficient solution for discrete RIS designs.

\appendices

\section{Proof of Proposition~\ref{prop:Bussgang_gain}} 
\label{appen:Buss_gain}
The Bussgang gain is defined as $\beta = \frac{\bbE[\mathcal{Q}(\breve{\phi}_m)\breve{\phi}_m^*]}{\bbE[|\breve{\phi}_m|^2]}$. Since $\breve{\phi}_m \sim \mathcal{CN}(0,1)$, the denominator is $\bbE[|\breve{\phi}_m|^2] = 1$.
Let the latent variable be expressed in polar coordinates as $\breve{\phi}_m = r e^{j\breve{\theta}}$, where the amplitude $r$ follows a Rayleigh distribution with $\bbE[r] = \frac{\sqrt{\pi}}{2}$, and the phase $\breve{\theta}$ is uniformly distributed in $[-\pi, \pi)$.
The quantizer $\mathcal{Q}(\cdot)$ maps the continuous phase $\breve{\theta}$ to the discrete phase ${\theta}$, such that the phase error $\delta = {\theta} - \breve{\theta}$ is uniformly distributed in $\left[-\frac{\pi}{L}, \frac{\pi}{L}\right)$.
The numerator can be computed as:
\begin{align}
     \nonumber
     \bbE[\mathcal{Q}(\breve{\phi}_m)\breve{\phi}_m^*] &= \bbE\left[ e^{j{\theta}} \cdot r e^{-j\breve{\theta}} \right] 
     \\
     &= \bbE[r] \cdot \bbE\left[ e^{-j(\breve{\theta} - {\theta})} \right] = \frac{\sqrt{\pi}}{2} \cdot \bbE\left[ e^{j\delta} \right].
\end{align}
The expectation of the phase error term is:
\begin{align}
    \bbE[e^{j\delta}] \!=\! \int_{-\frac{\pi}{L}}^{\frac{\pi}{L}} e^{j\delta} \frac{L}{2\pi} d\delta = \frac{L}{2\pi} \left[ \frac{e^{j\delta}}{j} \right]_{-\pi/L}^{\pi/L} 
    \!= \frac{L}{\pi} \sin\left(\frac{\pi}{L}\right)\!.
\end{align}
Multiplying the amplitude and phase expectations yields the generalized gain $\beta$ in \eqref{eq:Bussgang_gain}. 
This completes the proof.
\hfill $\blacksquare$

\section{Proof of Lemma~\ref{lem:NEP_RIS}}
\label{appen:GPI}
According to the stationary condition, the stationary points need to satisfy $\partial \cL(\bw)/\partial \bw^{\sf H} = 0$.
Thus, we take the partial derivative of $\cL(\bw)$ with respect to $\bw$ and set it to be zero. 
By using the derivative of the Rayleigh quotient form as
\begin{align}
    \label{eq:derivative_matrix}
    \frac{\partial \left( \frac{{\bw}^{\sf H} {\bJ}_k \bw}{{\bw}^{\sf H} {\bK}_k \bw}\right)}{\partial \bw^{\sf H}} = \left( \frac{\bw^{\sf H} {\bJ}_k \bw}{\bw^{\sf H} {\bK}_k \bw} \right) \left[\frac{{\bJ}_k \bw }{{\bw}^{\sf H} {\bJ}_k \bw} - \frac{{\bK}_k \bw }{{\bw}^{\sf H} {\bK}_k \bw} \right],
\end{align}

we can calculate the partial derivative of $\cL(\bw)$ in \eqref{eq:lamb_obj} as
\begin{align}
    \label{eq:partial_L1}
    \frac{\partial \cL (\bw)}{\partial \bw^{\sf H}} = \sum_{k=1}^K \frac{1}{\ln 2} \left( \frac{{\bJ}_k \bw }{{\bw}^{\sf H} {\bJ}_k \bw} - \frac{{\bK}_k \bw }{{\bw}^{\sf H} {\bK}_k \bw} \right).
\end{align}

Using \eqref{eq:partial_L1}, the stationary condition holds if 
\begin{align}
    \label{eq:kkt_cond_rearrange}
    \sum_{k=1}^{K}  \left(\frac{\bJ_k}{ {\bw}^{\sf H} \bJ_k \bw} \right)\bw
    =\sum_{k=1}^{K} \left(\frac{\bK_k}{ {\bw}^{\sf H} \bK_k \bw} \right)\bw.
\end{align}
The stationary condition can be regarded as the generalized eigenvalue problem with its corresponding matrices $\bar{\bJ}(\bw)$ and $\bar{\bK}(\bw)$ defined in \eqref{eq:CD_KKT}: $\bar{\bJ}(\bw) \bw = \lambda(\bw) \bar{\bK} (\bw) \bw.$
Here, $\bar{\bK}(\bw)$ is positive definite for $\sigma^2/P>0$ and positive $\lambda_{\sf 2}(\bw)$.
\hfill $\blacksquare$

\bibliographystyle{IEEEtran}
\bibliography{ref}

\end{document}